\documentclass[11pt]{article}
\usepackage{jheppub}

\usepackage{amsmath}
\usepackage{amssymb}
\usepackage{graphicx,color,slashed,braket}
\usepackage{bbm}
\usepackage{ifpdf}
\usepackage{comment}
\usepackage{soul}

\usepackage{booktabs}
\usepackage{multirow}
\usepackage{makecell}
\usepackage{float}
\usepackage{verbatim}
\usepackage{caption}
\usepackage{cancel}
\usepackage{enumerate}
\usepackage{tcolorbox}
\usepackage[font=small,labelfont=bf]{caption}
\usepackage[symbol]{footmisc}
\usepackage{color}
\definecolor{dark-gray}{gray}{0.20}
\definecolor{gray}{gray}{0.30}
\definecolor{light-gray}{gray}{0.80}
\definecolor{dark-red}{rgb}{0.7,0,0}
\definecolor{dark-green}{rgb}{0.1,0.4,0}
\definecolor{dark-blue}{rgb}{0.3,0.3,0.7}
\definecolor{light-blue}{rgb}{0.8,0.8,1}

\usepackage{pgfplots}
\pgfplotsset{width=10cm,compat=1.9}
\usepgfplotslibrary{external}
  
\usepackage{hyperref}
\usepackage{setspace}
\hypersetup{
	colorlinks=true,
	linkcolor=dark-blue,
	citecolor=dark-red,
	urlcolor=dark-green,
	linktoc=page
}
\newcommand{\be}{\begin{equation}}
\newcommand{\ee}{\end{equation}}

\def\be{\begin{equation}}
\def\ee{\end{equation}}
\def\bea{\begin{eqnarray}}
\def\eea{\end{eqnarray}}

\newcommand{\e}{\mathrm{e}}

\newcommand{\dd}{\mathrm{d}}

\newcommand{\rr}{\hat{r}}
\newcommand{\ttt}{\hat{t}}

\newcommand{\vol}{\text{vol}}

\newcommand{\lfour}{\ell_{\text{Pl},4}}
\newcommand{\lfive}{\ell_{\text{Pl},5}}

\allowdisplaybreaks

\usepackage[backgroundcolor=white!95!red, linecolor=cyan, bordercolor=cyan, textsize=footnotesize]{todonotes}

\title{Charged Nariai black holes on the dark bubble}
\preprint{UUITP-13/24}

\author{Ulf Danielsson \& Vincent Van Hemelryck}
\affiliation{Department of Physics and Astronomy, Uppsala University, Box 516, SE-75120, Uppsala, Sweden}

\emailAdd{ulf.danielsson@physics.uu.se}
\emailAdd{vincent.vanhemelryck@physics.uu.se}

\notoc

\abstract{In this paper, we realise the charged Nariai black hole on a braneworld from a nucleated bubble in AdS$_5$, known as the dark bubble model.
Geometrically, the black hole takes the form of a cylindrical spacetime pulling on the dark bubble. This is realised by a brane embedding in an AdS$_5$ black string background. Identifying the brane with a D3-brane in string theory allows us to determine a relation between the fine structure constant and the string coupling, $\alpha_\text{EM} = \frac{3}{2} g_s$, which was previously obtained for a microscopic black hole. We also speculate on the consequences for the Festina Lente bound and neutrino masses.}

\begin{document}

\maketitle

\newpage
\tableofcontents

\renewcommand{\thefootnote}{\arabic{footnote}}
\section{Introduction}

The construction of four-dimensional de Sitter vacua from string compactifications remains a tremendous challenge in string phenomenology \cite{Danielsson:2018ztv,Obied:2018sgi,Ooguri:2018wrx, Bedroya:2019snp}. Additionally, string models with phases of variable cosmic acceleration have been under pressure as well, see refs. \cite{Hebecker:2019csg,Cicoli:2021fsd, Cicoli:2021skd,Rudelius:2021azq,Rudelius:2022gbz, Andriot:2022xjh,Shiu:2023nph, Shiu:2023fhb, Hebecker:2023qke,Freigang:2023ogu}, however see refs. \cite{Andriot:2023wvg,Andriot:2024jsh} for studies with open universes. Therefore, the fate of cosmic acceleration in string compactifications remains an important topic of study in the Swampland program. All these investigations start from string compactifications down to a 4d effective field theory, but contrarily, in the dark bubble scenario, this starting point is left behind.

According to the dark bubble model, the universe is riding an expanding bubble in AdS$_5$, with the radial direction along the throat \cite{Banerjee:2018qey}. The bubble is formed through a phase transition, where a bubble of true vacuum nucleates. Its wall consists of a brane with a tension slightly less than the critical value, with the difference setting the value of the positive cosmological constant. The model has been studied and developed in a series of papers, see refs. \cite{Banerjee:2018qey,Banerjee:2019fzz,Banerjee:2020wix,Banerjee:2020wov, Banerjee:2021qei,Danielsson:2022fhd,Danielsson:2022lsl,Danielsson:2023alz,Basile:2023tvh,Banerjee:2023uto}, with the earlier results reviewed in refs. \cite{Banerjee:2021yrb,Banerjee:2022ree} and related studies in \cite{Berglund:2022qsb,Bandos:2023yyo,Koga:2019yzj,Koga:2020jok,Koga:2022opd,Basile:2020mpt,Basile:2021vxh}. In \cite{Danielsson:2022lsl,Danielsson:2023alz} the dark bubble model was embedded into a specific string theory setting, where the unique hierarchy of scales of the model was explored. In standard compactifications and RS-brane worlds, the 5d Planck length is always larger than the 4d one, provided that the size of the extra dimension, and the AdS length scale, are larger than the higher dimensional Planck length. The dark bubble, which has a brane with an inside and an outside, rather than the two insides of the RS-brane, is fundamentally different. What matters is the difference in AdS scale between the inside and the outside, which leads to a {\it smaller} 5d Planck length than the 4d one. This leads to several unique phenomenological features - including no need for scale separation in the higher dimensions. 

As the 4d spacetime is defined on a shell rather than in a vacuum of the theory, the standard rules of effective field theory do not directly apply to the 4d setup. For this reason, it is rather non-trivial to study the physics on the dark bubble, such as particle physics, gravitational waves, gauge theories or even black holes. Nevertheless, the dark bubble model needs to incorporate the matter fields we are familiar with in 4d. In \cite{Danielsson:2022fhd} it was shown how 4d gravitational waves are induced from 5d ones. The analysis took the expansion of the universe into account, and provided a non-trivial test of the model, demonstrating the intricate interplay between the 5d bulk, the embedding of the brane, and the induced metric in the 4d spacetime. In \cite{Basile:2023tvh}, a further step was taken, where electromagnetism was identified with the gauge field in the DBI-action of a D3-brane on which the universe is riding. To achieve this, it was crucial to include the presence of the $H= \dd B$ three-form field of string theory. As was shown in \cite{Basile:2023tvh}, the electromagnetic field strength in 4d sources an $H$-field in the bulk, which then backreacts on the 5d geometry. This affects the induced 4d metric such that the 4d gravitational backreaction is accounted for. The analysis there was made in an expanding universe, and is, again, a very non-trivial result, confirming the consistency of the dark bubble model.

In this paper, we consider another case that further tests the dark bubble model: a (charged) black hole. We will restrict ourselves to the simplest setup, a charged Nariai black hole as large as the cosmological horizon\footnote{While the Nariai solution is not relevant for a direct realistic application in cosmology, its embedding serves as an important check on the theoretical consistency of the dark bubble model.} \cite{Nariai:1999,Romans:1991nq}. Such black holes have been studied in the context of the Swampland paradigm \cite{Montero:2019ekk}, where it has been argued that their existence leads to new constraints on particle masses beyond those of the weak gravity conjecture. In this paper, we show that the dark bubble scenario can successfully incorporate a Nariai black hole. This specific construction serves as a stepping stone for realising more general black holes on the dark bubble. The analysis provides yet another explicit example of how the dark bubble reproduces 4d Einstein gravity without localising it on the brane, a characteristic of the RS model. Instead, the bulk plays an important role in the construction, and the failure to localise yields well-controlled corrections that become important only at stringy densities, \cite{Basile:2023tvh}. 

Furthermore, we find a relation between the fine structure constant and the string coupling, which is in agreement with the results of ref. \cite{Danielsson:2023alz}. There, using an explicit string theory embedding of a small, extremal black hole of unit charge, it was shown that the fine structure constant $\alpha_\text{EM}$ is fixed by the string coupling $g_s$ through
\begin{equation}
    \alpha_\text{EM} = \frac{3}{2} g_s \, .
\end{equation}
Remarkably, the same relation can be derived from the embedding of the macroscopic Nariai black hole.

The outline of the paper is as follows. In section \ref{sec:dark_bubble_scenario} we review some technical aspects of the dark bubble model. In section \ref{sec:RN-dS} we review the 4d Nariai solution. In section \ref{sec:5d BS} we study a 5d black string solution, which we use in section \ref{sec:Glueing} to uplift the Nariai solution. In section \ref{sec:alpha_gs} we match the fine structure coupling to the string coupling, and in section \ref{sec:discussion} we end with some conclusions. 

\section{The dark bubble scenario}\label{sec:dark_bubble_scenario}
In the dark bubble scenario, a bubble with an AdS$_5$ space inside is joined together with another AdS$_5$ spacetime on the outside by a brane. The dynamics of the brane are governed by two sets of key equations, i.e. the Israel junction conditions and the Gauss-Codazzi equations. The junction conditions are simply the 5d Einstein equations in the presence of a thin brane, relating the extrinsic curvatures on the in- and outside with the stress tensor on the brane. The Gauss-Codazzi equation relates the 5d bulk curvatures and the extrinsic curvatures of the embedding, to the intrinsic 4d curvatures of the brane. We discuss the junction conditions at length in section \ref{sec:junction_conditions} and refer the reader to its discussion there.

However, it is interesting to emphasise that these junction conditions and Gauss-Codazzi equations can be combined into the 4d Einstein equations, as was done in \cite{Banerjee:2019fzz}:
\begin{equation} \label{eq: 4d Einstein}
    G_{ab}^{(4)} = \kappa_4^2\left( (-\sigma _c h_{ab} - S_{ab} )+\frac{1}{2 \kappa_5^2}\left[\left(\frac{\mathcal{J}^+_{ab}}{k_+}- \frac{\mathcal{J}^-_{ab}}{k_-}\right) - \frac{1}{2}\left(\frac{\mathcal{J}^+}{k_+}- \frac{\mathcal{J}^-}{k_-}\right)h_{ab}\right] \right) = \kappa_4^2 T_{ab},
\end{equation}
where 
\begin{equation}
    \mathcal{J}_{ab} = R^{(5)}_{\alpha \beta \gamma \delta}e^{\alpha}_{\; c} e^{\beta}_{\; a}e^{\gamma}_{\;d} e^{\delta}_{\;b} h^{cd} \, ,
\end{equation}
and $e^{\alpha}_{\;b}$ are the tangent vectors to the brane for every $b$. The ``$-$'' and ``$+$'' correspond to the contributions from the in- and outside spacetimes respectively. The 4d Newton's constant is related to the 5d one through
\begin{equation}
    \kappa_4^2 = \frac{2 k_- k_+}{k_- - k_+} \kappa_5^2 \, ,
\end{equation}
where $\kappa_d^2 = 8 \pi G_d = \ell_{\text{Pl},d}^{d-2}$.\footnote{Here, we see the cause of the inverted hierarchy mentioned in the introduction. To get the result for an RS brane we need to replace $k_ - - k_+ \rightarrow k_- + k_+$. Usually, the scales are also chosen to be equal, i.e. $k_-=k_+=k$. In this way, the RS construction is very similar to a compactification with an extra dimension of size $R_\text{AdS} = 1/k$. In the case of the dark bubble, tuning $k_+$ to be just slightly smaller than $k_-$, creates an inverted hierarchy with $\lfive \ll \lfour$. }  We have also defined the critical tension
\begin{equation}
    \sigma _c = \frac{3}{\kappa_5^2} (k_- - k_+) \, .
\end{equation}
We see that the effective energy-momentum tensor has two contributions. The direct contribution from the brane itself, together with any matter that it carries in the form of $S_{ab}$, and a geometric contribution from the bulk curvature through the tensor $\mathcal{J}_{ab}$. The expression for the Einstein equation is not exact but has corrections at second order in $S_{ab}$ (but no higher orders), see ref. \cite{Banerjee:2019fzz} for more details. These corrections are small as long as the net contribution to the 4d Einstein equations does not reach stringy densities \cite{Basile:2023tvh}.

An important property, peculiar to the dark bubble, is that $S_{ab}$ seemingly contributes with the wrong sign to the 4d energy density. This is the minus sign in front of $S_{ab}$ in (\ref{eq: 4d Einstein}). In the case of a pure brane, we simply have $S_{ab}= -\sigma h_{ab}$, which is why a larger value of the brane density $\sigma$ {\it reduces} the effective cosmological constant given by $\Lambda_4 = \sigma _c- \sigma$.\footnote{Note that we are using $-+++$ signature so that $h_{tt}<0$ has $S_{tt}>0$ (and $S^t_t<0$) for a positive energy density.} For the dark bubble to nucleate, its tension needs to be smaller than the critical value $\sigma_c$, guaranteeing a positive cosmological constant.

To get the full energy-momentum tensor $T_{ab}$ in the presence of matter on the brane, one must take into account its backreaction on the bulk, captured by  $\mathcal{J}_{ab}$. This gives rise to an effective total energy density in 4d that is equal to the physical one. It is this phenomenon that has been exhibited in earlier work mentioned in the introduction. The choice of the bulk solution is part of setting up the appropriate initial conditions for the physical situation of interest. It is a non-trivial result of ref. \cite{Basile:2023tvh} that this can be done for electromagnetic waves such that the subsequent time evolution is that of general relativity coupled to Maxwell theory. A crucial aspect is how the electromagnetic field on the brane sources a string $B$-field in the bulk. The backreaction of this field on the bulk metric is what allows the tensor $\mathcal{J}_{ab}$ to give rise to the desired energy-momentum tensor $T_{ab}$.

Following ref. \cite{Basile:2023tvh}, we note that the 4d Einstein tensor, and the electromagnetic energy-momentum tensor on the brane, are both covariantly conserved. The first by construction, and the latter through its equations of motion. This is what is needed for any consistent coupling of matter to gravity. It then follows, by construction, that the extra piece in the 4d Einstein equations that changes the sign of the energy-momentum tensor, also is covariantly conserved. Hence, there is no mixing between the terms, and the Einstein equations remain true for all times, up to corrections that are relevant only at stringy energy densities.

The goal of this paper is to incorporate the charged Nariai black hole into the dark bubble scenario, which we review in the next section.

\section{Reissner-Nordstr\"om-de Sitter black holes and the Nariai limit}\label{sec:RN-dS}

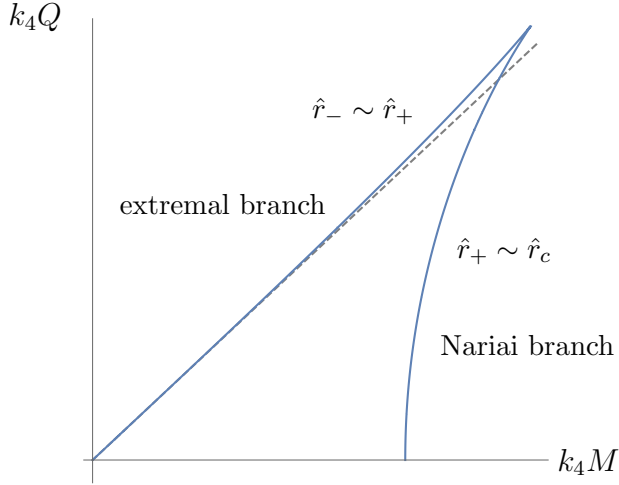
\begin{figure}[ht]
\centering
	\begin{tikzpicture}
    	\node[anchor=south west,inner sep=0] (image) at (0,0) {\includegraphics[width=0.4\textwidth]{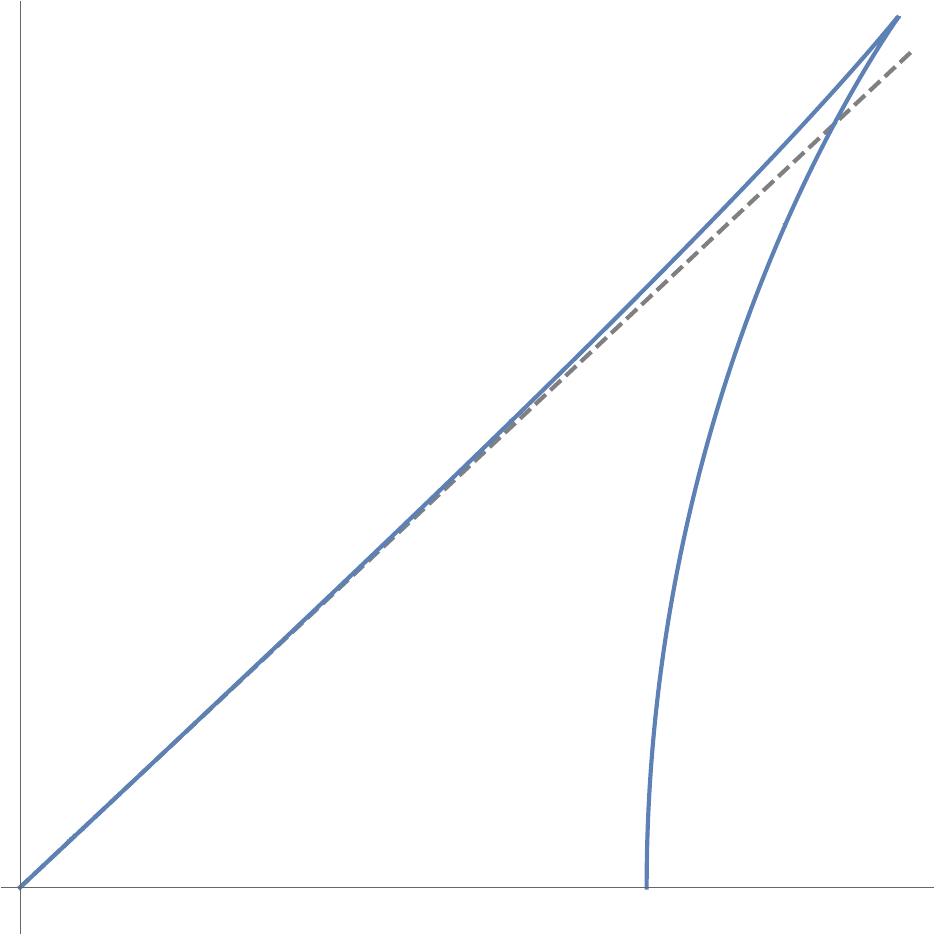}};
    \begin{scope}[
        x={(image.south east)},
        y={(image.north west)}
    ]
        \node [black] at (-0.1,1) {\large $k_4 Q$};
        \node [black] at (1.08,0.05) {\large $k_4 M$};
        \node [black] at (0.6,0.8) {\normalsize $\rr_- \sim \rr_+$};
        \node [black] at (0.9,0.5) {\normalsize $\rr_+ \sim \rr_c$};
        \node [black] at (0.30,0.6) {\normalsize extremal branch};
        \node [black] at (0.95,0.3) {\normalsize Nariai branch};
    \end{scope}
\end{tikzpicture}
\caption{Only black holes with charge and mass under the curve are allowed. The top blue line corresponds to extremal black holes where inner and outer black hole horizons meet, and is referred to as the `extremal branch'. The vertical blue line represents the black holes in the Nariai limit, where the outer horizon and cosmological horizon meet, which is referred to as the `Nariai branch'.The dashed line represents $Q=M$ and illustrates that the extremality bound for RN-dS black holes deviates from the one for RN black holes in Minkowski space.}
\label{fig:sharkfin}
\end{figure}

In this section, we briefly review Reissner-Nordstr\"om-de Sitter (RN-dS) black holes and the Nariai limit in four dimensions. These have been receiving increased attention recently and led to the Festina Lente bound in \cite{Montero:2019ekk}. These black holes are solutions of Einstein-Maxwell theory given by the familiar action
\begin{equation}
    S = \frac{1}{2\lfour^2} \int \dd^4 x \sqrt{-g_4} \left( R_4 - \frac{1}{4}F_{\mu \nu}F^{\mu \nu} - 2 \Lambda_4 \right)\,,
\end{equation}
with $\Lambda_4 = + 3 k_4^2$.
For finding black hole solutions, we choose the metric to take the usual expression
\begin{equation}
    \dd s_4^2 = -f(\rr) \dd \ttt^2 + \frac{\dd \rr^2}{f(\rr)} + \rr^2 \dd \Omega_2\,.
\end{equation}
The Maxwell equations then allow for a non-trivial electric field with the familiar form:
\begin{equation} \label{eq: const F}
    F = \frac{2 Q}{ \rr^2} \dd \ttt \wedge \dd \rr \, .
\end{equation}
Finally, the Einstein equations are solved for the following function $f$:
\begin{align}
\label{eq:RNdS_metric_function_QM}
    f(\rr) &= 1 - \frac{2 M}{\rr}+ \frac{Q^2}{\rr^2} - k_4^2 \rr^2\\
\label{eq:RNdS_metric_function_rprm}
    &= -\frac{k_4^2}{\rr^2}(\rr+ \rr_- + \rr_+ + \rr_c)(\rr-\rr_-)(\rr-\rr_+) (\rr-\rr_c).
\end{align}
The first formulation \eqref{eq:RNdS_metric_function_QM} contains 3 parameters, where $k_4$ is the inverse dS$_2$ radius and $M$ and $Q$ are the mass and charge parameters, respectively, both with dimension length. They are related to the physical mass $m$ and charge $Q_\text{EM}$ as follows:
\begin{equation}
\label{eq:physical_QM}
    M = \frac{\hbar c \lfour^2 }{8\pi} m\,, \qquad
    Q^2=\frac{Q_\text{EM}^2 \lfour^2}{32 \pi^2 \epsilon_0 \hbar c}.
\end{equation}
The second formulation \eqref{eq:RNdS_metric_function_rprm} contains 4 parameters, the three radii $\rr_+, \rr_-, \rr_c$ and $k_4$. Without loss of generality, we take $\rr_- \leq \rr_+ \leq \rr_c$, with $\rr_-$ ($\rr_+$) representing the inner (outer) horizon radius of the black hole and $\rr_c$ the cosmological horizon radius. These radii must satisfy one constraint, being:
\begin{equation}
    k_4^2\left(\rr_-^2 + \rr_+^2 + \rr_c^2 + \rr_-\rr_+ + \rr_-\rr_c +\rr_+\rr_c \right) = 1,
\end{equation}
which is solved for
\begin{equation}
\label{eq:BH_r_constraint}
    \rr_-= \frac{1}{2} \left(-(\rr_+ + \rr_c) + \sqrt{4 k_4^{-2} - 3( \rr_+^2 + \rr_c^2) - 2 \rr_+ \rr_c}\right).
\end{equation}
Relating the radii to the charge and mass parameters, we find
\begin{equation}
\label{eq:QM_rprmrc}
    Q^2 = k_4^2\: \rr_- \rr_+ \rr_c \left(\rr_- + \rr_+ + \rr_c \right), \qquad 2M = k_4^2 (\rr_- + \rr_+)(\rr_- + \rr_c)(\rr_+ + \rr_c) .
\end{equation}
The constraint \eqref{eq:BH_r_constraint} and the fact that $\rr_- \leq \rr_+ \leq \rr_c$, do not allow for every combination of charge and mass. For black holes in Minkowski space, this leads to the extremality bound $Q \leq M$. The constraint comes from the fact that the inner horizon cannot cross the outer horizon, i.e. $\rr_- \ngeq \rr_+$, preventing the development of a naked singularity. For the RN-dS black hole, the same reasoning goes true, although the extremality bound $Q \leq M$ is slightly altered. Additionally, the outer horizon of the black hole is not allowed to be bigger than the cosmological horizon. As the three radii are related through eq. \eqref{eq:BH_r_constraint}, these constraints can be summarised into one inequality, which in terms of the charge and mass takes the following form \cite{Romans:1991nq}:
\begin{equation}
    27 (k_4 M)^4-(k_4 M)^2 \left(36 (k_4 Q)^2+1\right)+(k_4 Q)^2\left(4 (k_4 Q)^2+1\right)^2 > 0.
\end{equation}
Both of these constraints can be visualised in the phase diagram of RN-dS black holes, depicted in figure \ref{fig:sharkfin}.

In the Nariai limit, one lets the cosmological horizon and outer black hole horizon come close together such that they become indistinguishable for observers. However, they cannot become strictly equal, and therefore it is useful to expand the metric for an observer that feels no force in between those two horizons, located at $\rr=\rr_g$ with $f'(\rr_g)=0$. The Nariai limit can then be understood by bringing the two horizons close to this static observer, although not touching.
In principle, one can find the solution to the equation $f'(\rr_g)=0$ in the form $\rr_g = \rr_g(\rr_+,\rr_c)$. However, the expression becomes rather lengthy and it is not necessary for our discussion, so we do not write it here.
So in what follows, we want to consider the limit $\rr_+ \rightarrow \rr_g \leftarrow \rr_c$, see also refs. \cite{Hawking:1995ap,Bousso:1996au}.
It proves to be useful to perform a coordinate transformation as follows:
\begin{equation}
    \rr \to \sqrt{|f(\rr_g)|}\: r + \rr_g, \qquad \ttt \to \sqrt{|f(\rr_g)|}\: t.
\end{equation}
With this, the 4d metric becomes
\begin{equation}
     \dd s_4^2 = -\frac{f(r)}{|f(\rr_g)|} \dd t^2 + \frac{|f(\rr_g)|}{f(r)}\dd r^2 + \left(\sqrt{|f(\rr_g)|} \;r + \rr_g\right)^2 \dd \Omega_2 .
\end{equation}
Now we take the Nariai limit $\rr_+ \rightarrow \rr_g \leftarrow \rr_c $.
For that, it is very useful to expand the functions $f(r)/|f(\rr_g)|$ in terms of $r$.
However, evaluating everything in terms of $\rr_g(\rr_+,\rr_c)$, which we did not write, and then taking the limit $\rr_+\to \rr_c$, is rather complicated. Instead, it turns out to be more useful to introduce a small parameter $\epsilon$ as follows:
\begin{equation}
    \rr_g = \rr_c (1- \epsilon), \qquad \rr_+ = \rr_c (1- 2 \epsilon + O(\epsilon^2)).
\end{equation}
Notice that the parameterisation above respects $f'(\rr_g)=0$ or equivalently $\rr_g=\rr_g(\rr_c,\rr_+)$, up to linear order in $\epsilon$.
Now, we can perform the whole calculation including up to first order in $\epsilon$ and then take $\epsilon \to 0$ in the Nariai limit.
Let us now perform a series expansion of the metric in small $r$ and up to leading order in $\epsilon$.
To do that, we note that
\begin{equation}
    \frac{\dd}{\dd r} = \sqrt{f(\rr_g)}\frac{\dd}{\dd \rr} .
\end{equation}
The metric functions become:
\begin{align}
\label{eq:near_Nariai_functions}
    \frac{f(r)}{|f(\rr_g)|}& & =& \:1 - \left[\left(6 k_4^2 - \frac{1}{\rr_c^2}\right) - \frac{2\epsilon}{\rr_c^2}\right] r^2 -\left[\sqrt{6 k_4^2 - \frac{1}{\rr_c^2}}(1-4 k_4^2 \rr_c^2) \frac{2\epsilon}{\rr_c^2}\right] r^3 + O(r^4,\epsilon^2)\\
    \rr^2& & =& \:\rr_c^2\left(1 -2 \epsilon + \left[2 \epsilon \sqrt{6 k_4^2 - \frac{1}{\rr_c^2}}\right]r + O(r^3,\epsilon^2)  \right)
\end{align}
In the Nariai limit, i.e. $\epsilon \to 0$, we recover the dS$_2 \times S^2$ metric:
\begin{equation}
\label{eq:metric_dS2xS2}
    \dd s_4^2 = - \left( 1 - (k_2r)^2\right) \dd t^2 + \frac{\dd r^2}{\left( 1 - (k_2 r)^2\right)} + \rr_c^2 \dd \Omega_2^2  ,
\end{equation}
with
\begin{equation}
\label{eq:k2_L_def}
    k_2^2 = 6 k_4^2 - \frac{1}{\rr_c^2}  .
\end{equation}
One can express these quantities in terms of the charge parameter $Q$:
\begin{equation}
\label{eq:k2_L_solutions}
    k_2^2= \frac{6 k_4^2\sqrt{1- 12 k_4^2Q^2}}{1+\sqrt{1-12 k_4^2 Q^2} }, \qquad \rr_c^2 = \frac{1+\sqrt{1-12 k_4^2 Q^2} }{6 k_4^2}  .
\end{equation}
It is important to note that, apart from being a near-horizon limit of the RN-dS solution, the Nariai metric is also a genuine solution of the Einstein-Maxwell equations with a constant electric field. Indeed, with the metric \eqref{eq:metric_dS2xS2} and constant electric field like the one in \eqref{eq: const F} (with $\rr=\rr_c$), the Einstein equations reduce to 
\begin{align}
    &- \frac{1}{\rr_c^2} = -\frac{Q^2}{\rr_c^4} -3 k_4^2\\
    &-k_2^2 = +\frac{Q^2}{ \rr_c^4} - 3 k_4^2\,,
\end{align}
which have the same solutions as in eq. \eqref{eq:k2_L_solutions}.
Furthermore, the Nariai background is also a solution of ``Einstein-Born-Infeld'' theory \cite{Cai:2004eh}, where one replaces the Maxwell action of electromagnetism with the Born-Infeld action, with a constant electric field. Additionally, the role of dilatonic couplings has also been studied for the charged Nariai black hole setup \cite{Bousso:1996pn}.

As explained in the introduction, the goal of this paper is to embed the Nariai black hole into the dark bubble.
Let us therefore end this section with some expectations on how the 5d embedding should look. In the dark bubble model, any 4d energy density on the brane bends it outwards. The first example was discussed in \cite{Banerjee:2019fzz}, where point particles in 4d correspond to strings that pull on the brane in the extra dimension. Additionally, in ref. \cite{Basile:2023tvh} it was shown that in the presence of an electromagnetic field on the brane, there also exists a $B$-field that pulls the brane outwards.

\begin{figure}[h]
\centering
    \includegraphics[width=12cm]{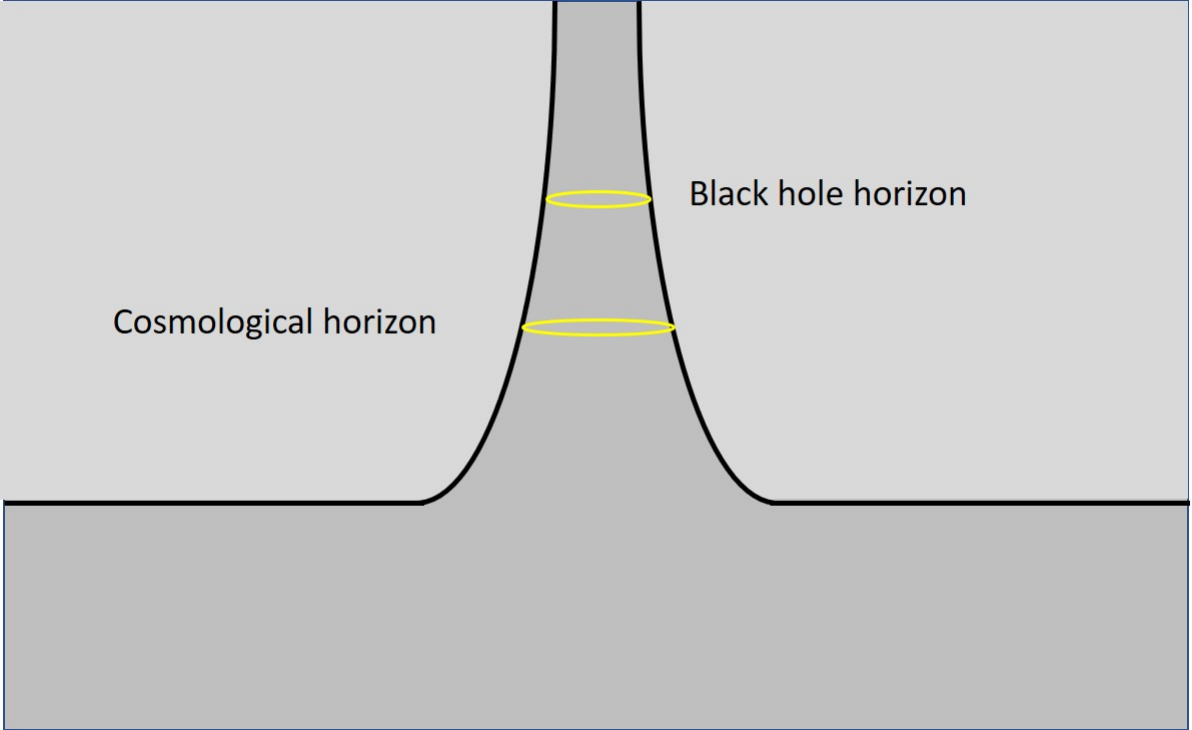} 
\caption{A black hole on top of a dark bubble. In the Nariai limit, the structure close to the horizons becomes a straight cylinder.}
\label{fig:NariaiThroat}
\end{figure}

It is therefore expected that other 4d energy sources, like black holes, also bend the brane.  Pictorially, one can think of a black hole as a very thick string pulling on the brane, and we expect a throat-like structure as in figure \ref{fig:NariaiThroat}. For an observer outside the black hole, there are two horizons. The outer one is the cosmological one, while the inner one is that of the black hole. The mass of the black hole is so large that the brane starts to bend upwards already far outside of the cosmological horizon centred on the black hole.
As explained above, in the Nariai regime, the black hole horizon and cosmological horizon do not coincide but are close together such that they become indistinguishable. In figure \ref{fig:NariaiThroat}, this corresponds to a strongly warped regime where the segment of the brane in between the two horizons becomes a cylinder rather than a cone.

In the next section, we explain in detail how a Nariai black hole on the brane can be understood in a 5d embedding. The upshot is that it can be seen as a perturbed version of a ``black string'' background, which was discussed in \cite{Bernamonti:2007bu}.

\section{The 5d black string}\label{sec:5d BS}
In this section, we set up the five-dimensional backgrounds that allow us to embed a brane with the Nariai geometry. For that, we need a five-dimensional asymptotically AdS spacetime with the proper topology. It turns out that black string solutions in AdS$_5$ do precisely that. 
Black strings in such spacetimes have been extensively studied, see 
refs. \cite{Klemm:2000nj, Bernamonti:2007bu, Brihaye:2007ju,Brihaye:2009dm,Harmark:2007md} for a select overview.
The background that is suitable for our purpose is related to the ``magnetically charged black string'' in AdS$_5$ that was described in ref. \cite{Bernamonti:2007bu}. In the remainder of this section, we introduce a modified version of this background, allowing for a dS$_2$ subspace.

In particular, we start from five-dimensional Einstein gravity coupled to a three-form field strength $H$ with the following action:
\begin{equation}
    S = \frac{1}{2\lfive^3} \int \dd^5 x \sqrt{-g_5} \left(R_5 - 3 \Lambda_5 - \frac{1}{2}|H|^2 \right)\,,
\end{equation}
where $\Lambda_5 = - 4k^2$ is the 5d cosmological constant with $k$ the inverse AdS$_5$ length, and $H$ a three-form field strength with two-form gauge potential $B$.\footnote{In a string theory context, the cosmological constant arises from moduli stabilisation. Notice also that we have omitted a dilatonic coupling to $H$. We have justified this in Appendix \ref{app:dilatonic_coupling}, where it is shown that a non-trivial dilaton can be neglected in the approximation scheme used in this paper.}
For the metric, we take a modified black string Ansatz that makes the dS$_2$ structure of the Nariai black hole that we are aiming for more apparent:
\begin{equation}\label{eq:5d_metric}
    \dd s_5^2 = - f(u)\left(1- (k_2 r)^2\right) \dd t^2 + \frac{\tilde f(u)}{\left(1- (k_2 r)^2\right)} \dd r^2 + u^2 \dd \Omega_2 + \frac{ \dd u^2}{w(u)}\,,
\end{equation}
where $k_2$ represents the inverse dS$_2$ radius. When it is put to zero, the metric corresponds to the Ansatz of \cite{Bernamonti:2007bu, Mann:2006yi} again.
We also turn on one non-vanishing, electric component of the $H$-field, which we parametrise by 
\begin{equation}
\label{eq:H_Ansatz}
    H = q \:  \frac{\sqrt{f(u)\tilde f(u)}}{u^2 \sqrt{w(u)}} \: \dd t \wedge \dd r \wedge \dd u\,,
\end{equation}
where $q$ is a constant. With this rather particular choice, the field strength $H$ satisfies
\begin{equation}
    \star H = q \: \vol (S^2), \qquad   \dd \star H =0\,,
\end{equation}
hence it automatically solves the equation of motion for $H$ and $q$ represents an electric charge for the field strength $H$.\footnote{In ref. \cite{Bernamonti:2007bu}, the authors work with a two-form field strength that we have Hodge dualised into the three-form field strength $H$. In their prescription, the constant $q$ is a magnetic charge of their two-form field strength.}
As usual, the Einstein equations are
\begin{equation}
    R_{MN} - \frac{1}{2}R g_{MN} + \frac{3}{2}\Lambda_5 = \frac{1}{2}\left(|H|_{MN}^2 - \frac{1}{2}g_{MN}|H|^2\right),
\end{equation}
This gives five non-independent equations to solve (the $tt$-, $rr$-, $\theta \theta$-, $uu$- and $ru$-components). They are the following:
\begin{align}
    &\left(\frac{\tilde{f}'}{4 \tilde{f}}+\frac{1}{u}\right) w'+w \left(-\frac{\tilde{f}'^2}{4 \tilde{f}^2}+\frac{\frac{\tilde{f}''}{2}+\frac{\tilde{f}'}{u}}{\tilde{f}}+\frac{1}{u^2}\right)-6 k^2+\frac{q^2}{u^4}-\frac{1}{u^2} = 0\\
    &\left(\frac{f'}{4 f}+\frac{1}{u}\right) w'+w \left(-\frac{f'^2}{4 f^2}+\frac{\frac{f''}{2}+\frac{f'}{u}}{f}+\frac{1}{u^2}\right)-6 k^2+\frac{q^2}{u^4}-\frac{1}{u^2} = 0\\
    \notag
    &\frac{w}{2}\left(\frac{ f''}{f}+\frac{ \tilde{f}''}{ \tilde{f}}\right)+\frac{w f' \tilde{f}'}{4 f \tilde{f}}+\left(\frac{w'}{4}+\frac{w}{2u}\right)\left(\frac{f'}{ f}+\frac{\tilde{f}'}{ \tilde{f}}\right)- \frac{w}{4}\left(\frac{f'^2}{ f^2}+\frac{ \tilde{f}'^2}{ \tilde{f}^2}\right)-\frac{k_2^2}{\tilde{f}}\\
    &-6 k^2-\frac{q^2}{u^4}+\frac{w'}{2 u} = 0\\
    &\frac{w f' \tilde{f}'}{4 f \tilde{f}}+\frac{w}{u}\left(\frac{f'}{ f}+\frac{\tilde{f}'}{ \tilde{f}}\right)-\frac{k_2^2}{\tilde{f}}-6 k^2+\frac{q^2}{u^4}+\frac{w}{u^2}-\frac{1}{u^2} =0\\
\label{eq:EinsteinEq_ru}
    &k_2^2\left(\frac{\tilde{f}'}{\tilde{f}}- \frac{f'}{f}\right) = 0\,.
\end{align}
It is important to note that all the derivatives appearing above are with respect to the $u$-coordinate and that the differential equations involve no $r$-dependence.
As we mentioned before, when $k_2$ vanishes, one recovers the Einstein equations for the charged black string in AdS$_5$ \cite{Bernamonti:2007bu}. Fully analytic solutions have only been found for special values of the charge $q$, see e.g. \cite{Bernamonti:2007bu, Chamseddine:1999xk,Klemm:2000nj}, and when $k^2 q^2 = 1/12$ the solution is even supersymmetric \cite{Chamseddine:1999xk,Klemm:2000nj}. Nevertheless, the Einstein equations can be solved both for small and large values of $k u$ for any charge $q$. We are interested in the latter case, where the functions $f$, $\tilde{f}$ and $w$ allow Fefferman-Graham expansions of the following form \cite{Bernamonti:2007bu, Mann:2006yi}\footnote{As explained in appendix \ref{app:dilatonic_coupling}, if a dilatonic coupling to the $H$-field were to be included, its equation of motion would impose its non-constant part to be much more subleading in the large $(ku)$-limit than the order we are considering. Therefore, it is justified to have omitted it.}:
\begin{align}
\label{eq:f_0th_order_exp}
        &f^{(0)}(u) = (k u)^2 + \frac{1}{2} + \frac{1}{12}(1- 12 (k q)^2)\frac{\log (k u)}{(k u)^2} + \frac{c_{t}}{(k u)^2} + O\left(\frac{\log(ku)}{(ku)^{4}} \right)\\
        &\tilde{f}^{(0)}(u) = (k u)^2 + \frac{1}{2} + \frac{1}{12}(1- 12 (k q)^2)\frac{\log (k u)}{(k u)^2} + \frac{c_{r}}{(k u)^2} + O\left(\frac{\log(ku)}{(ku)^{4}} \right)\\
\label{eq:w_0th_order_exp}
        &w^{(0)}(u) = (k u)^2 + \frac{2}{3} + \frac{1}{6}(1- 12 (k q)^2)\frac{\log (k u)}{(k u)^2} + \frac{c_{t}+c_{r} + (k q)^2/3}{(k_\pm u)^2} + O\left(\frac{\log(ku)}{(ku)^{4}} \right)
\end{align}
The constants $c_t$ and $c_r$ determine the mass and tension of the black string and can be determined in terms of the horizon radius by matching these expressions to the near-horizon expansions of the black string solutions, as was done in ref. \cite{Bernamonti:2007bu}.

When $k_2$ is turned on, the Einstein equations receive additional terms and hence the solutions change. First, eq. \eqref{eq:EinsteinEq_ru} informs us that $f = c' \tilde{f}$ and we can take $c'=1$ as it can be absorbed by redefinition of the time coordinate $t$. This implies that we need to take  $c_t = c_r$, and such 5d solutions turn out to be extremal at zeroth order in $k_2$.
We then solve the system of differential equations perturbatively when $k_2^2/k^2$ becomes small. We can write the expansions
\begin{align}
    &f(u) = f^{(0)}(u) + \frac{k_2^2}{k^2} f^{(1)}(u)  + O(k_2^4/k^4)\\
    &w(u) = w^{(0)}(u) + \frac{k_2^2}{k^2} w^{(1)}(u)  + O(k_2^4/k^4)\,.
\end{align}
When plugging these into the Einstein equations, we find the following Fefferman-Graham expressions of the first-order corrections:
\begin{align}
\label{eq:f_1st_order_exp}
    &f^{(1)}(u) = a_{1} (k u)^2 + \frac{1}{2}(a_{1} - 1)  + \frac{1}{12}a_{1} (1- 12 (k q)^2)\frac{\log (k u)}{(k u)^2} + \frac{c_{1}}{(k u)^2} + O\left(\frac{\log(ku)}{(ku)^{4}} \right)\\
\label{eq:w_1st_order_exp}
    &w^{(1)}(u) = -\frac{1}{3} + \frac{-2 (-c_{1}+c_{r}a_{1})}{(k u)^2} + O\left(\frac{\log(ku)}{(ku)^{4}} \right)\,.
\end{align}
This introduces two new parameters, $a_1$ and $c_1$, which are not fixed by the Einstein equations. One could try to see whether these parameters are constrained when matching these expansions with small $(ku)$-expansions as in \cite{Bernamonti:2007bu} and how the mass and tension of the black string background are affected, but we postpone this for the future.

\section{Glueing two black string solutions together}\label{sec:Glueing}

\subsection{Solving the junction conditions}\label{sec:junction_conditions}

At this stage, we are ready to glue two different black string spacetimes together. They are joined together by sharing a boundary, which corresponds to a (dynamical) shell. In our setting, the shell is provided by a three-brane that is embedded at a fixed location in the $u$-direction, i.e. at $ u = a$ with $a$ constant.
The prescription for glueing spacetimes was originally developed by Israel in the form of junction conditions \cite{Israel:1966rt}. The first condition requires the induced metric on the shell (three-brane) to be continuous:
\begin{equation}
\label{eq:1nd_junction_condition}
    h_{mn}^- = h_{mn}^+\,
\end{equation}
where $h_{mn}$ is the induced metric at $u=a$ and the ``$-$'' and ``$+$'' refer to the spacetime on the inside and outside respectively.
The second junction condition requires that the extrinsic curvature must be discontinuous across the shell. The jump in the extrinsic curvature is provided by the stress tensor of the shell, formulated in the second junction condition:
\begin{equation}
\label{eq:2nd_junction_condition}
    \Delta K_m^n - \Delta K h_m^n = -\lfive^3 S_m^n,
\end{equation}
where $S_{mn}$ is the stress tensor on the brane and the $\Delta$ refers to the difference of the quantities between inside and outside spacetimes, i.e.
\begin{equation}
    \Delta K_m^n = (K_-)_m^n - (K_+)_m^n, \qquad \Delta K = K_- - K_+.
\end{equation}
Before discussing the extrinsic curvature contributions, let us first take a look at the stress tensor on the brane. We use a three-brane with a world-volume field strength $\mathcal{F}$, described by a DBI action of the following form:
\begin{equation}
    S_\text{DBI} = -\sigma \int \dd^4 x\sqrt{-\det({h_{ab}} +\tau \mathcal{F}_{ab})}\,.
\end{equation}
The field strength $\mathcal{F}$ is a combination of a worldvolume U(1) gauge field $A$ with field strength $F$ and the pull-back of the $B$-field on the brane:
\begin{equation}
    \tau \mathcal{F} = \tau F_{ab} + B_{ab}\,,
\end{equation}
where $H= \dd B$ locally. With a stringy embedding in mind, we think of the brane as a D3-brane and we have $\tau = 2 \pi \alpha'$, while the tension of the D3-brane is $\sigma = \frac{1}{(2 \pi )^3 \alpha '^2 g_s}- \delta \sigma$. The correction $\delta \sigma$ reduces the tension so that $\sigma < \sigma_c$, and we get a positive cosmological constant. Its presence was argued by referring to the weak gravity conjecture in \cite{Danielsson:2022lsl}, and it was explicitly calculated in \cite{Danielsson:2023alz} using stringy $R^2$-corrections to the brane action.
We must also allow for a constant electric field on the brane. Due to the coupling to the $B$-field, such an electric field sources a jump in the $H$-field across the brane. As we have already imposed the expression of the $H$-field in the previous section, we can use the $H$-equation of motion with the three-brane source to evaluate the electric field and hence the stress tensor in terms of a jump in the $H$-field. The details are worked out in appendix \ref{app:Smn_brane}, and when we evaluate the stress tensor in the large $u = a$ limit, we find
\begin{align}
\label{eq:stress_tensor_brane_tt_rr}
    &S_{t}^t = S_r^r = - \sigma - \frac{1}{2\lfour^2}\frac{E^2}{2}   = - \sigma - \frac{1}{8} \frac{(\Delta q)^2}{\kappa_5^4}\frac{1}{\sigma}\frac{1}{a^4} \\
\label{eq:stress_tensor_brane_thetatheta_phiphi}
    &S_{\theta}^\theta = S_\phi^\phi = - \sigma + \frac{1}{2\lfour^2}\frac{E^2}{2} = - \sigma + \frac{1}{8} \frac{(\Delta q)^2}{\kappa_5^4}\frac{1}{\sigma}\frac{1}{a^4}.
\end{align}
Here $\Delta q = q_+ - q_-$ corresponds to the difference in 5d charges on the in- (``$-$'') and outside (``$+$'') as defined in eq. \eqref{eq:H_Ansatz}.
Now we turn back to the geometrical part. We write the metric on the out- and inside as follows:
\begin{equation}
    \dd s_5^2 = - f_{\pm}(u)\left(1- (k_{2\pm} r_\pm)^2\right) \dd t_\pm^2 + \frac{f_\pm(u)}{\left(1- (k_{2\pm} r)^2\right)} \dd r_\pm^2 + u^2 \dd \Omega_2 + \frac{\dd u^2}{w_\pm(u)},
\end{equation}
Note that we allowed for different time and spatial coordinates on the in- and outside with $t_\pm$ and $r_\pm$, and for different inverse de Sitter radii $k_{2\pm}$. Additionally, the functions $f_\pm(u)$ and $w_\pm(u)$ have the same functional form as in eqns. \eqref{eq:f_0th_order_exp}-\eqref{eq:w_0th_order_exp} and \eqref{eq:f_1st_order_exp}-\eqref{eq:w_1st_order_exp}, but we allow all the parameters to be different. This amounts to replacing $k \to k_\pm$, $c_r \to c_{r\pm}$, $q \to q_\pm$, $a_{1}\to a_{1\pm}$ and $c_{1}\to c_{1\pm}$. 

Now we have everything to solve the junction conditions, for two spacetimes joined together by a three-brane at $u=a$. The first junction condition requires us to rescale the time and radial coordinates as follows, defining new coordinates $t$ and $r$:
\begin{equation}\label{eq:coord_rescalings}
     \sqrt{f_+(a)} \: t_+ = \sqrt{f_-(a)} \: t_- = t, \qquad \sqrt{f_+(a)} \: r_+ = \sqrt{f_-(a)} \: r_- = r
\end{equation}
Additionally, as the induced metric has to be continuous, we have to impose that
\begin{equation}
\label{eq:k2_def}
    k_2 \equiv \frac{k_{2+}}{\sqrt{f_+(a)}} = \frac{k_{2-}}{\sqrt{f_-(a)}} \,,
\end{equation}
where the new $k_2$ is now the physical inverse dS$_2$ radius that an observer on the brane would measure.
With all this, induced metric becomes indeed of the dS$_2 \times S^2$ type:
\begin{equation}
    h_{mn}^{\pm} \dd x^m \dd x^n = - \left(1- (k_{2} r)^2\right) \dd t^2 + \frac{1}{\left(1- (k_{2} r)^2\right)} \dd r^2 + a^2 \dd \Omega_2
\end{equation}
Because the solutions are constructed in the large $(k_\pm a)$- and small $(k_{2\pm}/k_\pm)$-limit, we infer from eq. \eqref{eq:f_0th_order_exp} and \eqref{eq:k2_def} that $k_2$ must satisfy
\begin{equation}
\label{eq:hierarchy}
    k_2 (k_\pm a) \sim k_{2\pm} \Rightarrow \frac{k_2}{k_\pm} \sim \frac{k_{2\pm}}{k_\pm} \frac{1}{k_{\pm}a}  \ll \frac{1}{k_{\pm}a} \ll 1 \,.
\end{equation}
Next, we compute the extrinsic curvature combinations appearing in the junction condition \eqref{eq:2nd_junction_condition}. The $tt$- and $rr$-components are the same, the $\theta\theta$- and $\phi\phi$-component likewise, which results in only two independent components, which are
\begin{align}
\label{eq:extrinsic_tt}
    & \Delta K_t^t - \Delta K h_t^t = \frac{2}{a}\left(\sqrt{w_-(a)} - \sqrt{w_+(a)}\right) + \frac{1}{2} \left(\frac{\sqrt{w_-(a)} \: f_-'(a)}{f_-(a)} -\frac{\sqrt{w_+(a)} \: f_+'(a)}{f_+(a)} \right) \\
\label{eq:extrinsic_thetatheta}
    & \Delta K_\theta^\theta - \Delta K h_\theta^\theta = \frac{1}{a}\left(\sqrt{w_-(a)} - \sqrt{w_+(a)}\right) + \left(\frac{\sqrt{w_-(a)} \: f_-'(a)}{f_-(a)} -\frac{\sqrt{w_+(a)} \: f_+'(a)}{f_+(a)} \right)
\end{align}
These expressions are valid for any functions $f_{\pm}(a)$ and $w_\pm(a)$. 

We are now ready to glue two particular geometries together.
For the inside, we choose a geometry described by \eqref{eq:f_0th_order_exp}-\eqref{eq:w_0th_order_exp} and \eqref{eq:f_1st_order_exp}-\eqref{eq:w_1st_order_exp}, with appropriate constants $c_r$ etc. such that the 5d geometry in the $k_2 \to 0$ limit is a 5d black string solution as in ref. \cite{Bernamonti:2007bu}.

On the outside, we allow us a bit more freedom. We still choose a similar black string solution, but we allow ourselves to choose the constant $c_r$ as we please. We can do so for the following reason: the values of $c_{r}$ that were calculated in ref. \cite{Bernamonti:2007bu} were obtained by requiring that the large $(k u)$-expansion of the functions can be glued onto the expressions of the near-horizon expansion, where the horizon was located at $k u_h \lesssim O(1)$. However, in our setup on the outside, we do not need such requirements. What happens for small $k u$, all the way down to the would-be horizon, is taken care of by the geometry on the inside of the bubble. This allows for more freedom in choosing the parameters. A similar situation occurred in ref. \cite{Danielsson:2022odq}, where over-extremal Reisner-Nordström geometries were constructed, also by glueing two AdS spacetimes together with a spherical shell. A naked singularity did not occur, as the inside of the shell did carry no charge. 

With these comments, we are now ready to compute the extrinsic curvature contributions to the junction conditions.
As we discussed before, the functions $f_\pm$ and $w_\pm$ are defined up to first order in $k_{2\pm}^2/k^2$ as in \eqref{eq:f_1st_order_exp}-\eqref{eq:w_1st_order_exp}. These should now be expressed in terms of the physical $k_2$ through the first junction condition \eqref{eq:k2_def}. Since we have constructed the solutions perturbatively in $k_{2\pm}$, we can identify the physical $k_2$ in a similar manner, by which we mean that
\begin{equation}
    k_{2\pm}^2 = k_2^2 f_{\pm}(a) = k_2^2 \left(f_\pm^{(0)}(a)+ \frac{k_{2\pm}^2}{k_\pm^2}f_\pm^{(1)}(a) + \cdots \right)
\end{equation}
and hence our perturbative expansion allows us to drop the second term and write 
\begin{equation}
    k_{2\pm}^2 = k_2^2 f_\pm^{(0)}(a). 
\end{equation}
Additionally, we note that the functions $w_\pm$ and $f_\pm$ are defined up to order $(k_\pm a)^{-2}$, but we expand the extrinsic curvature contributions \eqref{eq:extrinsic_tt}-\eqref{eq:extrinsic_thetatheta} up to order $(k_\pm a)^{-4}$. We should do this because the functions $w_\pm$ and $f_\pm$ are leading with $(k_{\pm} a)^2$ and approximated up to $(k_{\pm} a)^{-2}$, but by taking the root in \eqref{eq:extrinsic_tt}-\eqref{eq:extrinsic_thetatheta} lowers that with one power, and dividing by $a$ or taking a derivative lowers it with an additional one. Hence the combinations are leading by $(k_\pm a)^0$ and are truncated after $(k_\pm a)^{-4}$. Finally, we also allow us to throw away terms of the order $(k_2^2/k_\pm^2) /(k_\pm a)^2$ because these are much smaller than $1/(k_\pm a)^4$ by eq. \eqref{eq:hierarchy}.
With all these truncations, the extrinsic curvature contributions become
\begin{align}
\notag
    \Delta K_t^t - \Delta K h_t^t =& 3(k_- - k_+) -\frac{k_- - k_+}{2k_- k_+}\frac{1}{a^2} + \left(\frac{c_{r-}-\frac{1}{24}}{k_-^3}-\frac{c_{r+}-\frac{1}{24}}{k_+^3}\right)\frac{1}{a^4} \\
    &+\left(\frac{(1-12 k_-^2 q_-^2)\log (k_- a)}{1 k_-^3} - \frac{(1-12 k_+^2 q_+^2)\log (k_+ a)}{12 k_+^3}\right)\frac{1}{a^4}\\
\notag
    &+ k_2^2 \Biggl[ (a_{1-}k_- -a_{1+}k_+) a^2 + \left(\frac{3 - 2 a_{1+}}{6 k_+}-\frac{3 - 2 a_{1-}}{6 k_-}\right)\Biggr]\\
\notag
    \Delta K_\theta^\theta - \Delta K h_\theta^\theta =& 3(k_- - k_+)-\left(\frac{c_{r-}-\frac{1}{24}-\frac{1-12 k_-^2 q_-^2}{24}}{k_-^3}-\frac{c_{r+}-\frac{1}{24}-\frac{1-12 k_+^2 q_+^2}{24}}{k_+^3}\right)\frac{1}{a^4}\\
\notag
    &-\left(\frac{(1-12 k_-^2 q_-^2)\log (k_- a)}{12 k_-^3} - \frac{(1-12 k_+^2 q_+^2)\log (k_+ a)}{12 k_+^3}\right)\frac{1}{a^4}\\
    &+ k_2^2 \Biggl[ 2(a_{1-}k_- -a_{1+}k_+) a^2 + \left(\frac{3 - 4 a_{1+}}{6 k_+}-\frac{3 - 4 a_{1-}}{6 k_-}\right)\Biggr]
\end{align}
We notice that the logarithmic dependence on $a$ in both of the extrinsic curvature contributions are the same, and that the junction conditions become insensitive to the constants $c_{1\pm}$ due to our approximations.
Equating these to the stress tensor contributions \eqref{eq:stress_tensor_brane_tt_rr}-\eqref{eq:stress_tensor_brane_thetatheta_phiphi} and rearranging the terms, we find that the junction conditions become
\begin{align}
\label{eq:junction1}
    &-\frac{1}{a^2} + k_2^2\left(A(k_\pm, a_{1\pm})\; a^2 + B(k_\pm, a_{1\pm}) \right) = - \frac{Q_t^2}{a^4} - \frac{\Xi(a)}{a} - 3k_4^2\\
\label{eq:junction2}
       &-k_2^2\left(2A(k_\pm, a_{1\pm})\; a^2 + 2B(k_\pm, a_{1\pm}) + 1 \right) = + \frac{Q_\theta^2}{a^4} +\frac{\Xi(a)}{a^4}- 3k_4^2,
\end{align}
where we have introduced the following definitions:
\begin{align}
    &A(k_\pm, a_{1\pm}) = 2\frac{k_- k_+}{k_- - k_+}(a_{1-}k_- -a_{1+}k_+)\\
    &B(k_\pm, a_{1\pm}) = 2\frac{k_- k_+}{k_- - k_+} \left(\frac{3 - 2 a_{1+}}{6 k_+}-\frac{3 - 2 a_{1-}}{6 k_-}\right)\\
    \label{eq: Xi}
    &\Xi(a) = \frac{2 k_- k_+}{k_- - k_+}\left(\frac{(1-12 k_-^2 q_-^2)\log (k_- a)}{12 k_-^3} - \frac{(1-12 k_+^2 q_+^2)\log (k_+ a)}{12 k_+^3}\right)\\
      \label{eq:Qt}
    &Q_t^2 = \frac{2 k_- k_+}{k_- - k_+}\left(-\frac{1}{8} \frac{(\Delta q)^2}{\lfive^3}\frac{1}{\sigma} + \frac{c_{r-}-\frac{1}{24}}{k_-^3}-\frac{c_{r+}-\frac{1}{24}}{k_+^3}\right)\\
    \label{eq:Qtheta}
    &Q_\theta^2 = Q_t^2 + \frac{2 k_- k_+}{k_- - k_+}\left(\frac{1-12k_+^2 q_+^2}{24 k_+^2}-\frac{1-12k_-^2 q_-^2}{24 k_-^2}\right)\\
    &3 k_4^2 = -\frac{2 k_- k_+}{k_- - k_+} \left(\lfive^3 \sigma - 3(k_- - k_+) \right)\,.
\end{align}

\subsection{Matching with 4d Nariai}

Ideally, we want the junction conditions \eqref{eq:junction1}-\eqref{eq:junction2} to take the form of the 4d Einstein equations in the usual 4d Nariai setting:
\begin{align}
\label{eq:EinsteinEq_4d_tt}
    -\frac{1}{a^2} &= -\frac{Q^2}{ a^4}-3 k_4^2\\
\label{eq:EinsteinEq_4d_thetatheta}
    -k_2^2 &= +\frac{Q^2}{ a^4}-3 k_4^2  .
\end{align}
This is not quite the case yet. For instance, we want no term proportional to $k_2^2$ in \eqref{eq:junction1}, meaning that $A(k_\pm, a_{1\pm})=0$ and $B(k_\pm, a_{1\pm})=0$. Additionally, we want no term proportional to $k_2^2 a^2$ in \eqref{eq:junction2}, the only term involving $k_2^2$ being exactly proportional to $-1$, which is accounted for by the same requirements. These are solved by
\begin{equation}
    a_{1\pm} = \frac{3 k_\mp}{2(k_- + k_+)}.
\end{equation}
With these choices, there are a few other terms that we do not want to be there, e.g. the ones with a logarithmic dependence in $a$. There are a few ways to tackle this problem.

\subsubsection*{A special case}

If one tunes the 5d charges so they obey
\begin{equation}
    k_+^2 q_+^2 = \frac{k_-^3 - k_+^3(1-12k_-^2 q_-^2) }{12 k_-^3} \, ,
\end{equation}
we find that the function $\Xi(a)$ becomes constant and that the two charge parameters are the same:
\begin{equation}
    \Xi(a) = \Xi = \frac{2 k_- k_+}{k_- - k_+} \frac{(1-12 k_-^2 q_-^2)\log\left( \frac{k_-}{k_+}\right)}{12 k_-^3}, \qquad Q_t^2 = Q_\theta^2\,.
\end{equation}
Now the two junction conditions \eqref{eq:junction1}-\eqref{eq:junction2} nicely become the 4d Einstein equations of the Nariai spacetime with the 4d charge satisfying:
\begin{equation}
\label{eq:Qsquared_from_5d}
    Q^2 = Q_t^2 + \Xi =  \frac{2 k_- k_+}{k_- - k_+} \left[ -\frac{1}{8} \frac{(\Delta q)^2}{\lfive^3}\frac{1}{\sigma} + \frac{c_{r-}-\frac{1}{24}}{k_-^3}-\frac{c_{r+}-\frac{1}{24}}{k_+^3} + \frac{(1-12 k_-^2 q_-^2)\log\left( \frac{k_-}{k_+}\right)}{12 k_-^3}\right]
\end{equation}
In these expressions, it is useful to identify the 4d Planck scale using the standard dark bubble relation
\begin{equation}
\label{eq:lfour_ito_lfive}
    \lfour^2=\frac{2 k_- k_+ }{k_- - k_+} \lfive^3.
\end{equation}
Per usual, the Nariai system is solved for $a$ and $k_2$ in terms of $Q$ and $k_4$. 
We find that
\begin{equation}
    k_2^2 a^2 = \sqrt{1- 12 k_4^2Q^2} \qquad k_4^2 = \frac{1+ k_2^2 a^2}{6a^2}
\end{equation}
and for the tension, we have that
\begin{equation}
    \lfive^3 \sigma = 3(k_- - k_+) - \frac{1}{2}\left(\frac{1}{k_+}-\frac{1}{k_-} \right)3k_4^2 = 3(k_- - k_+) \left(1 -\frac{k_4^2}{2k_- k_+}\right)
\end{equation}
We need to check as a consistency requirement that $(k_2 a)^2 \ll 1$ and $\sigma >0$, and preferably that $\lfive^3 \sigma - 3(k_- - k_+) \ll 1$. The first requirement stems from eq. \eqref{eq:hierarchy} and is accounted for whenever $1-12(k_4 Q)^2 \ll 1$. This means that the Nariai black holes we are describing, are near-extremal, and hence the chargeless-limit $k_4 Q\to 0$ cannot be taken.
Additionally, one sees that for the near-extremal limit, $ k_4^2 \approx 1/(6a^2)$. Hence $k_4^2/(2k_+k_-) \approx 1/(12 (k_+a)(k_-a)) \ll 1$ and the tension is indeed positive and close to the critical tension. The near-extremal Nariai limit can be obtained by an appropriate choice of $c_{r+}$. We want to stress again that, although $q_+$ and $c_{r+}$ are not completely independent in the ordinary black string setup, we can take them independent here as we do not need to meet the requirements that the solution on the outside extrapolates to a solution for small values of $a$ without naked singularities.

\subsubsection*{Unconstrained charges}

If the extra term on the right-hand side of eq. \eqref{eq:Qtheta}, as well as the term $\Xi$  in eq. \eqref{eq:junction1}, were subleading, we would not need to impose any further constraints on the charges. This can be achieved provided that $\Delta k = k_- - k_+ \ll k_-$, and by taking $q_-=0$, $k_+ q_+ \ll 1$. Having $k_- \sim k_+$ is natural in the stringy embeddings considered in \cite{Danielsson:2022lsl,Danielsson:2023alz}. As in the examples studied in \cite{Basile:2023tvh}, it is also natural to have $q_- =0$ and the exterior charged $q_+$ fully sourced by the presence of an electromagnetic field on the brane. As we verify later, we also find that, even though the 4d Nariai is near-extremal, the 5d uplift has $k_+ q_+ \ll 1$, which means that it is far from supersymmetric.

\section{Relating fine structure constant to the string coupling}\label{sec:alpha_gs}

In this section, we derive a relation between the gauge coupling of the electric field, i.e. the fine structure constant, and the string coupling. Before doing so, let us remind ourselves of some qualitative features of our setup. As we explained at the end of section \ref{sec:RN-dS}, it was shown in \cite{Banerjee:2019fzz} that fundamental strings pulling on the brane provide very small black holes on the brane, and we argued that larger black holes can be seen as thick strings. Those can be seen as multiple strings puffed up into a cylinder-like structure, where the strings dissolve into electric flux on the brane.
Let us tie this idea with the properties of the mass-charge diagram or sharkfin of 4d black holes in figure \ref{fig:sharkfin}. A black hole arising from one fundamental string pulling on the brane, which has unit values of charge and mass, sits in the lower-left corner of the diagram and corresponds to the smallest possible, extremal black hole. If several such strings are added on top of each other, one moves up along the extremal branch, where the inner and the outer horizons of the black hole coincide. This can be done up to the maximal Nariai value in the upper right corner of the diagram, where both black hole horizons coincide with the cosmological horizon.

There exists a very non-trivial way to test the dark bubble model of the Nariai black hole. In ref. \cite{Banerjee:2019fzz}, it was shown that a string pulling on the brane corresponds to a particle of mass $\tau_\text{ps} L$, where $\tau_\text{ps}$ is the effective tension of the string and $L=1/k$, the AdS radius.
The effective tension can be obtained by integrating the energy density over the $S^2$. The difference in tension (including the energy density of the electric flux) compared to the critical tension contributes to the energy density as the cosmological constant.

According to the picture painted above (i.e. seeing the extremal Nariai solution as a polarised collection of fundamental hanging strings), this should be true also for the thick string representing the extremal Nariai solution, implying that
\begin{equation}
\label{eq:tauL_is_M}
     \tau _{\rm Nariai} L =m \, .
\end{equation}
In the following, we show that this indeed holds, provided that the fine structure constant of electromagnetism is related to the string coupling, precisely as in ref. \cite{Danielsson:2023alz}.

To test this condition, let us first compare our charge $Q$ with its usual definition in Maxwell's theory in terms of the fine structure constant and fundamental charge. We make use of the assumptions at the end of the previous section and find using eq. \eqref{eq:Qt}
\begin{equation}
     Q^2 = \lfour^2 \left(-\frac{1}{8} \frac{q_+^2}{\lfive^6}\frac{1}{\sigma} + \frac{c_M}{\lfive^3 k^3}\right) \, ,
\end{equation}
where
\begin{equation}
    c_M=c_{r-}-c_{r+} \,.
\end{equation}
Here we only include leading contributions in $k_- - k_+$ and have put $k_- \sim k_+ \sim k$. We double-check our approximations at the end. Note that $c_M>0$ and $c_{r+} <0$. We first need to properly quantise the charges. To do this, we recall eq. \eqref{eq:physical_QM} and reinstate the speed of light and reduced Planck's constant:
\begin{equation}
    M = \frac{\hbar c \lfour^2}{8\pi} m\,, \qquad
    Q^2=\frac{Q_\text{EM}^2 \lfour^2}{32 \pi^2 \hbar \epsilon_0 c}=N^2 \alpha_\text{EM} \frac{\lfour^2}{8\pi} .
\end{equation}
We then take $Q_\text{EM}=N e$, where $e$ is the elementary charge, and use $\alpha_\text{EM}=\frac{e^2}{4 \pi \epsilon_0 \hbar c}$, the fine structure constant. As explained above, we can think of this 4d charge being supplemented by $N$ fundamental strings pulling on the brane. These strings also source the $H$-field on the outside, providing the 5d charge $q_+$.
Considering how a fundamental string sources the $H$-field (see Appendix \ref{app:F1_charge}), we find
\begin{equation}
    q_\text{F1}=\frac{\lfive^3}{4 \pi^2 \alpha '}\,,
\end{equation}
and hence we can take $q_+ = N q_\text{F1}$ (and $q_-=0$). All this implies that
\begin{equation}
   N^2 \alpha_\text{EM} \frac{\lfour^2}{8\pi} = -\frac{N^2 g_s \lfour^2}{16\pi}   + \frac{\lfour^2 c_M}{\lfive^3 k^3}\,,
\end{equation}
and we conclude that we must choose
\begin{equation}
    \frac{\lfour^2 c_M}{\lfive^3 k^3} = \frac{N^2 \lfour^2}{8\pi} \left( \alpha_\text{EM} + \frac{g_s}{2 }\right)\,.
\end{equation}
To further test eq. \eqref{eq:tauL_is_M}, let us consider an extremal Nariai black hole. Such black holes sit in the upper right corner of the charge-mass diagram in figure \ref{fig:sharkfin}, have a Mink$_2 \times S^2$ near-horizon geometry and correspond to taking $k_2 = 0$.
We now want to view the extremal Nariai black hole as a thick string pulling on the dark bubble.
To do this we must first identify the effective tension. This is obtained by integrating the correct energy density over the $S^2$. The total energy density as measured in 4d is given by $\frac{1}{\lfour^2 a^2}$. This is {\it not} the relevant quantity as observed from 5d. Several quantities contribute to the total energy density (or tension) and they can be read from how the 5d spacetime is curved. The strategy to find this energy density is to view one of the junction conditions as the energy conservation equation. Following \cite{Brown:1988kg}, we multiply the junction condition in \eqref{eq:2nd_junction_condition} by the factor $\left( \sqrt{w_-(a)} + \sqrt{w_+(a)} \right)/2$, evaluate the extrinsic curvature contributions and rearrange to get (to leading orders)
\begin{equation}
    -S^t_t \frac{\sqrt{w_-(a)} + \sqrt{w_+(a)}}{2}-\frac{3(k_-^2 - k_+^2)a}{2 \lfive^3} = \frac{c_M}{\lfive^3 k^2 a^3}
\end{equation}
Integrating over $S^2$ to get a tension, multiplies the expression with $4\pi a^2$. The first (positive) term on the left-hand side is the total energy density of the brane (expressed in 5d units), including the electromagnetic contribution, while the second (negative) term is due to the lower energy inside the bubble. The right-hand side is the total effective tension, $\tau_{\rm Nariai}$, as viewed from afar.
We find
\begin{equation}
\label{eq:mass_from_junction}
    m = \tau_{\rm Nariai} L = \frac{4 \pi L c_M }{\lfive^3 k^2 a} =  N^2 \left( \alpha_\text{EM} + \frac{g_s}{2 }\right) \frac{1}{2 a}
\end{equation}
We find the same formula by looking at the masses of the black string spacetimes, which were computed in ref. \cite{Bernamonti:2007bu}. The effective energy or mass of the string should be equal to the difference in masses of the out- and inside spacetimes as energy is conserved in the bubble nucleation process. From ref. \cite{Bernamonti:2007bu}, we see that the mass, adapted to our case, is
\begin{equation}
    m_{\pm} = \frac{4 \pi L }{ \lfive^3 k_\pm \sqrt{f_\pm(a)}} \left(\frac{1}{24} -  c_{r\pm} \right).
\end{equation}
We have an additional factor $\sqrt{f_\pm(a)}$ due to the coordinate rescalings that we have introduced in eq. \eqref{eq:coord_rescalings}. The effective mass of the brane to leading order in our approximations is
\begin{equation}
    m = m_+ - m_- = \frac{4\pi L c_M}{\lfive^3 k^2 a},
\end{equation}
which agrees perfectly with \eqref{eq:mass_from_junction}.

To proceed, we look at the extremal Nariai limit, for which 
\begin{equation}
    a=\frac{1}{\sqrt{6} k_4} \qquad
    Q = \frac{1}{\sqrt{12} k_4} \qquad
    \frac{\lfour^2}{8\pi} m = \frac{2 \sqrt{2}}{3} Q < Q \, ,
\end{equation}
where the third relation is obtained by putting $\rr_- = \rr_+ = \rr_c$ in eq. \eqref{eq:QM_rprmrc}. This leads to
\begin{equation}
     m = \tau _{\rm Nariai} L  = \frac{8\pi Q^2}{\lfour^2 \alpha_\text{EM}} \left( \alpha_\text{EM} + \frac{g_s}{2 }\right) \frac{1}{\sqrt{2} Q}  = \frac{3 m}{4 \alpha_\text{EM}} \left( \alpha_\text{EM} + \frac{g_s}{2 }\right)
\end{equation}
This equation is only consistent when we require that  $\alpha_\text{EM} = \frac{3 g_s}{2}$. What we see is that our identification of the 4d gauge field with the electromagnetic Maxwell field forces the string coupling $g_s$ to be directly related to the fine structure constant $\alpha_\text{EM}$ in 4d. 

This is {\it exactly} the same value obtained through the very different calculation of \cite{Danielsson:2023alz} using unit charged extremal black hole as a fundamental string pulling on the dark bubble. The calculation does not make use of any resolution of the string into a 4d structure as in the case of the cosmological Nariai black hole. Instead, it relies on the detailed microscopic stringy embedding into higher dimensions, in particular on an AdS$_5 \times S^5$ background \cite{Danielsson:2022lsl}. Let us review the calculation and compare.

The 4d extremal RN black hole with unit charge has the mass 
\begin{equation}
    m_\text{RN} = \sqrt{\frac{8\pi \alpha_\text{EM}}{\lfour^2}} \, ,
\end{equation}
while the effective 4d mass of the pulling string is given by
\begin{equation} \label{eq: mp}
    m_\text{ps}= \tau_\text{ps} L = \frac{L}{2 \pi \alpha'} = \sqrt{\frac{12\pi g_s}{ \lfour^2}} \, .
\end{equation}
Here we have used the relation
\begin{equation} \label{eq: comb}
    \frac{L^2 \lfour^2}{\alpha '^2}= 48 \pi^3 g_s \,.
\end{equation}
This can be obtained by combining different equations, such as the relation between the 10d and 5d Planck scale, as well as the expressions of the AdS$_5$ scale and $S^5$ radius in terms of the $F_5$-flux quantum $N_c$, which are
\begin{equation}
\label{eq:AdS5xS5_data}
    \lfive^3 = \frac{\kappa_{10}^2}{{\rm Vol}(S^5)} = 2^6 \pi^4 g_s^2 \frac{\alpha '^4}{L^5} \, , \qquad
    L^4 = 4 \pi g_s N_c \alpha '^2 \,.
\end{equation}
The final piece of information needed for eq. \eqref{eq: comb} is the junction condition \eqref{eq:lfour_ito_lfive}, which by eq. \eqref{eq:AdS5xS5_data}, is 
\begin{equation} 
    \lfour^2 = \frac{3 N_c}{L} \lfive^3 \, .
\end{equation}
The relation eq. \eqref{eq: comb} is also what guarantees that the tension of a pure D3-brane (ignoring the subleading correction that gives the positive cosmological constant) is {\it exactly} equal to the critical one $\sigma_c$. That is, 
\begin{equation}
    \frac{1}{(2\pi )^3 \alpha '^2 g_s}= \sigma _c =\frac{3 \Delta k}{ \lfive^3} =\frac{6}{L^2 \lfour^2} \, .
\end{equation}
Demanding $m_\text{RN}=m_\text{ps}$, we again find $ \alpha_\text{EM}= \frac{3}{2} g_s$.

What is remarkable for the extremal Nariai black hole, is that the derivation of the relation between the string coupling and the fine structure constant is done {\it without} needing details of a stringy embedding. Nevertheless, this was necessary in the analysis in ref. \cite{Danielsson:2023alz}, for a single string as reviewed above. 
This indicates that the dark bubble is a very robust construction. In fact, the logic can be turned around by using the dark bubble to derive the necessary properties of the fundamental objects representing, e.g., unit-charged extremal black holes. What we find is that strings in a background such as AdS$_5 \times S^5$ have exactly those properties.

\section{Discussion}\label{sec:discussion}

In this paper, we embedded the extremal Nariai black hole into the dark bubble scenario. We did so by joining two (modified) AdS$_5$ black string backgrounds together, accommodating the proper isometries and asymptotics of the Nariai solution. We interpreted this as the black hole representing a thick string pulling on the brane. Nevertheless, what happens for more general black holes remains an interesting question. Such black holes are expected to warp the brane more, and it would be interesting to see how the junction conditions can be solved for such backgrounds. See ref. \cite{Banerjee:2021qei}, for some related results.

Next, one could wonder about Gregory-Laflamme instabilities \cite{Gregory:1993vy} for the setup we consider. According to the correlated stability conjecture \cite{Gubser:2000ec,Gubser:2000mm}, a classical Gregory-Laflamme instability should correspond to a thermodynamic instability, i.e. when the specific heat becomes negative. This has been verified for many examples, although counterexamples have been found as well where the black strings have scalar hair, see e.g. refs. \cite{Friess:2005zp,Buchel:2010wk,Buchel:2011ra}. In \cite{Bernamonti:2007bu}, where no scalar hair is present and hence the conjecture is expected to hold, a critical value of the 5d charge was identified beyond which the specific heat is always positive. Below this value, three different branches open up of which two are stable. It turns out that the extremal black string solution we consider on the outside, i.e. with $c_r = c_t$, has zero Hawking temperature and belongs to a stable branch.
For the inside, where we are also interested in extremal background with vanishing 5d charge, it was shown in ref. \cite{Brihaye:2007ju} that the classical instability does arise for this regular background. A possible solution to this problem is to consider a non-vanishing 5d charge on the inside as well. Notice that all these observations are only studied for the case $k_2 =0$. It would be interesting to see how the situation is affected by non-vanishing $k_2$ in the future. Additionally, one should remember that these instabilities have been investigated for static backgrounds. In the full dark bubble setup that we have in mind, the black hole is more than the cylinder, as depicted in figure \ref{fig:NariaiThroat}. The black hole is riding a brane that is expanding. It would be interesting to see if there is a connection between this and classical instability. 

Next, we managed to relate the electromagnetic fine structure constant to the string coupling, in agreement with the calculation for an extremal black hole of unit charge, as in \cite{Danielsson:2023alz}. Remarkably, the same relation is obtained even though our calculation for the cosmological Nariai black hole does not make use of the microscopic relations between, e.g., the string scale and the higher dimensional scale $L$. This suggests an internal consistency of the dark bubble that is replicated by string theory. 

Finally, as we constructed Nariai geometries, it is only natural to consider the Festina Lente bound, discussed in refs. \cite{Montero:2019ekk, Montero:2021otb}, which states that any charged particle in a (quasi-) de Sitter background must obey
\begin{equation}
\label{eq:fl}
    m^4 \geq m_4^2 H^2 \sim 1/(N_c^{2} \lfour^4) \sim  1/L^4 ,
\end{equation}
where we have expressed the bound in quantities relevant to the dark bubble. The reason for this bound becomes apparent when the charge-mass diagram of figure \ref{fig:sharkfin} is studied. The weak gravity conjecture ensures that a charged black hole can be discharged by emitting particles of mass $Q>M$. However, when the mass of the particle becomes too small, the charged black holes close to the Nariai branch in figure \ref{fig:sharkfin}, could exit the allowed region, and collapse into a naked singularity. The Festina Lente bound prevents this from happening.

We believe that our embedding of the Nariai black hole into the dark bubble can be used to further examine this bound. As we see from (\ref{eq:fl}), the bound exactly coincides with the size of the AdS$_5$ scale and the size of the extra dimension. According to Festina Lente, all charged particles must have a mass at least set by this scale. This is particularly true for all fermions. As already speculated in \cite{Basile:2023tvh}, we expect that the embedding of the standard model into the dark bubble involves sets of branes, possibly separated on the internal five-dimensional manifold. Most of the fermions in the Standard Model, such as the electrons, should have masses which are given by short strings connecting separated branes. This is equivalent to a coupling to the Higgs field. The neutrinos, which happen to have masses of the order $1/L$, are a possible exception. Here, the masses could be due to a coupling between massless modes on the brane (the flavour eigenstates) to massive modes probing the bulk. It could even be that {\it all} fermions are of this form, with a minimum mass set by $1/L$. It is just that most fermions have further contributions to the mass from the coupling to the Higgs fields. This, then, would suggest an intriguing connection between the Festina Lente bound and the neutrino masses.

\section*{Acknowledgements}
We want to thank Daniel Panizo and Thomas Van Riet for interesting discussions.
VVH is supported by Olle Engkvists Stifelse.

\appendix
\section{The dilatonic coupling}\label{app:dilatonic_coupling}
In this section we write down the conventions for the scales and couplings we use, argue that the dilaton can be taken as a constant, and derive the energy-momentum tensor of the brane, which was used in section \ref{sec:Glueing}. To connect the black string backgrounds with string theory, we start from the 10d NSNS action in Einstein frame for the dilaton:
\begin{equation}
    S_\text{NSNS} = \frac{1}{2\kappa_{10}^2 g_s^2}\int \dd^{10} x\sqrt{-g_{10}} \left(R_{10} -\frac{1}{2}(\partial \phi)^2 - \frac{1}{2} \e^{-(\phi - \phi_0)}|H|^2\right)\,,
\end{equation}
with $H = \dd B$ and where $\e^{\phi_0}=g_s$. Notice that this has been obtained by Weyl rescaling the original string frame action with the dynamical part of the 10d dilaton only. When the action is reduced to 5d, where we allow only for an $H$-field along the 5d external spacetime, the reduced NSNS action retains the same form
\begin{equation}
\label{eq:NSNSaction_EF}
    S_\text{NSNS} = \frac{1}{2\kappa_{5}^2}\int \dd^5 x \sqrt{-g_{5}} \left(R_{5} -\frac{1}{2}(\partial \phi)^2 - \frac{1}{2} \e^{-(\phi - \phi_0)}|H|^2\right) \,,
\end{equation}
where the 5d Planck constant satisfies
\begin{equation}
    \kappa_5^2 = \text{Vol}_5 \;  g_s^2 \; \kappa_{10}^2
\end{equation}
Notice that a 5d cosmological constant can arise from moduli stabilisation from internal fluxes (RR and NSNS). We also remark that there is a dilatonic coupling to the $H$-field. We have not included it in the main text, as we assumed the dilaton to be constant, so the dilatonic coupling disappears. We justify this assumption here below. With the presence of a dynamical dilaton, we infer that its equation of motion is 
\begin{equation}
    \nabla^2 \phi + \frac{1}{2}\e^{-(\phi - \phi_0)} |H|^2 = \frac{1}{\sqrt{-g_5}} \partial_\mu \left( g^{\mu \nu} \sqrt{-g_5}\; \partial_\nu \phi \right)+ \frac{1}{2}\e^{-(\phi - \phi_0)} |H|^2 =0\,.
\end{equation}
We make the dilaton dependent on the $u$-coordinate, such that with the metric \eqref{eq:5d_metric} and $H$-field \eqref{eq:H_Ansatz}, this equation of motion can be written as
\begin{equation}
    \frac{\sqrt{w}}{f u^2} \: \left(f\sqrt{w}\; u^2 \: \frac{g'}{g} \right)'+ \frac{1}{2} \frac{g_s}{g} u^{-4} q^2 =0
\end{equation}
where $\e^{\phi(u)} = g(u)$. With the large $ku$-expansions \eqref{eq:f_0th_order_exp}-\eqref{eq:w_0th_order_exp} for $f$ and $w$, one can solve this for $g(u)$ and finds that the $ku$-dependence is subleading:
\begin{equation}
    g(u) = g_s + O\left(\frac{\log(ku)}{(ku)^{4}} \right)\,,
\end{equation}
which justifies taking the dilaton to be constant. If one wants to solve our setup beyond this order, it is necessary to keep dilatonic couplings in the 5d Einstein equations and solve for all the metric functions and dilaton accordingly, essentially solving for a dilatonic, magnetic AdS$_5$ black string. This is beyond the scope of this work.

\section{Energy-momentum tensor on the brane}\label{app:Smn_brane}
In this appendix, we derive the stress tensor on the brane. We do so by looking at the DBI action of a D3-brane in Einstein frame for the dilaton. It is given by 
\begin{equation}
    S_\text{DBI} = -\sigma \int \dd^4 x \sqrt{- \det\left(h_{ab} + \tau \mathcal{F}_{ab} \e^{-(\phi-\phi_0)/2} \right)}\,,
\end{equation}
with $\sigma = \mu_3/g_s = 1/((2\pi)^3 \alpha'^2 g_s)$, $\tau = 2\pi \alpha'$ and 
\begin{equation}
    \tau \mathcal{F}_{ab} = \tau F_{ab}+ B_{ab}\,.
\end{equation}
When using a constant dilaton, we have
\begin{equation}
    S_\text{DBI} = -\sigma \int \dd^4 x \sqrt{- \det\left(h_{ab} + \tau \mathcal{F}_{ab} \right)}\,.
\end{equation}
Now we discuss how this brane sources the $B$-field. Let us expand the square root such that the DBI action becomes:
\begin{equation}
    S_\text{DBI} = \int \dd^4 x \sqrt{-h_4} \left(-\sigma - \frac{\sigma \tau^2}{2} |\mathcal{F}|^2 \right)
\end{equation}
This is the Maxwell action where $\sigma \tau^2$ is the inverse gauge coupling squared.
This field strength sources the $H$-field. Indeed, we find that
\begin{equation}
    \frac{\delta S_\text{DBI}}{\delta B} = -  \sigma \tau \star_4 \mathcal{F} \wedge \delta(u-a) \dd u,
\end{equation}
On the other hand, we find from \eqref{eq:NSNSaction_EF} that
\begin{equation}
    \frac{\delta S_\text{NSNS}}{\delta B} = \frac{1}{2\kappa_5^2} \dd \star_5 H\,,
\end{equation}
so the equation of motion for $H$ is
\begin{equation}
    \frac{1}{2\kappa_5^2} \dd \star_5 H = \sigma \tau \star_4 \mathcal{F} \wedge \delta(u-a) \dd u\,.
\end{equation}
With an electric Ansatz $\mathcal{F} = E \dd t \wedge \dd r$ and using our expression for $H$, eq. \eqref{eq:H_Ansatz} in terms of the 5d magnetic charges, we find that this equation is solved when the electric field satisfies:
\begin{equation}
    E = \frac{\Delta q}{2\kappa_5^2}\frac{1}{ \sigma \tau}\frac{1}{a^2}. 
\end{equation}
At last, we compute the stress tensor on the brane, which is given by
\begin{align}
    &S_{t}^t = S_r^r = - \sigma - \sigma \tau^2 \frac{E^2}{2}   = - \sigma - \frac{1}{2} \frac{(\Delta q)^2}{(2\kappa_5^2)^2}\frac{1}{\sigma}\frac{1}{a^4} \\
    &S_{\theta}^\theta = S_\phi^\phi = - \sigma + \sigma \tau^2 \frac{E^2}{2} = - \sigma + \frac{1}{2} \frac{(\Delta q)^2}{(2\kappa_5^2)^2}\frac{1}{ \sigma}\frac{1}{a^4} \,.
\end{align}
Notice that instead of expanding the square root in the DBI action, one could also choose not to do so. The $H$ equation of motion then imposes that the electric field $E$ must take the standard Born-Infeld form. Fortunately, expanding that expression for large $a$ yields the same electric field at leading order and hence the same stress tensor as was found above.

\section{Fundamental string charge} \label{app:F1_charge}
Finally, we briefly comment on how a fundamental string sources the $B$-field, to derive the fundamental charge in 5d. A fundamental string couples to the $B$-field as follows:
\begin{equation}
    S_\text{WZ, F1} = -\frac{1}{2\pi \alpha'} \int B\,.
\end{equation}
It therefore sources the $B$-field in its equation of motion by
\begin{equation}
   \frac{1}{2 \kappa_5^2} \dd \star_5 H - \frac{1}{2\pi \alpha'} \delta_\text{F1} = 0.
\end{equation}
Integrating this and using Gauss' law, we obtain the fundamental charge as 
\begin{equation}
    q_\text{F1} = \frac{ \kappa_5^2}{4\pi^2 \alpha'} \,.
\end{equation}

\bibliographystyle{JHEP}
\bibliography{refs.bib}

\providecommand{\href}[2]{#2}\begingroup\raggedright\begin{thebibliography}{10}

\bibitem{Danielsson:2018ztv}
U.~H. Danielsson and T.~Van~Riet, \emph{{What if string theory has no de Sitter
  vacua?}}, \href{https://doi.org/10.1142/S0218271818300070}{\emph{Int. J. Mod.
  Phys. D} {\bfseries 27} (2018) 1830007}
  [\href{https://arxiv.org/abs/1804.01120}{{\ttfamily 1804.01120}}].

\bibitem{Obied:2018sgi}
G.~Obied, H.~Ooguri, L.~Spodyneiko and C.~Vafa, \emph{{De Sitter Space and the
  Swampland}},  \href{https://arxiv.org/abs/1806.08362}{{\ttfamily
  1806.08362}}.

\bibitem{Ooguri:2018wrx}
H.~Ooguri, E.~Palti, G.~Shiu and C.~Vafa, \emph{{Distance and de Sitter
  Conjectures on the Swampland}},
  \href{https://doi.org/10.1016/j.physletb.2018.11.018}{\emph{Phys. Lett. B}
  {\bfseries 788} (2019) 180}
  [\href{https://arxiv.org/abs/1810.05506}{{\ttfamily 1810.05506}}].

\bibitem{Bedroya:2019snp}
A.~Bedroya and C.~Vafa, \emph{{Trans-Planckian Censorship and the Swampland}},
  \href{https://doi.org/10.1007/JHEP09(2020)123}{\emph{JHEP} {\bfseries 09}
  (2020) 123} [\href{https://arxiv.org/abs/1909.11063}{{\ttfamily
  1909.11063}}].

\bibitem{Hebecker:2019csg}
A.~Hebecker, T.~Skrzypek and M.~Wittner, \emph{{The $F$-term Problem and other
  Challenges of Stringy Quintessence}},
  \href{https://doi.org/10.1007/JHEP11(2019)134}{\emph{JHEP} {\bfseries 11}
  (2019) 134} [\href{https://arxiv.org/abs/1909.08625}{{\ttfamily
  1909.08625}}].

\bibitem{Cicoli:2021fsd}
M.~Cicoli, F.~Cunillera, A.~Padilla and F.~G. Pedro, \emph{{Quintessence and
  the Swampland: The Parametrically Controlled Regime of Moduli Space}},
  \href{https://doi.org/10.1002/prop.202200009}{\emph{Fortsch. Phys.}
  {\bfseries 70} (2022) 2200009}
  [\href{https://arxiv.org/abs/2112.10779}{{\ttfamily 2112.10779}}].

\bibitem{Cicoli:2021skd}
M.~Cicoli, F.~Cunillera, A.~Padilla and F.~G. Pedro, \emph{{Quintessence and
  the Swampland: The Numerically Controlled Regime of Moduli Space}},
  \href{https://doi.org/10.1002/prop.202200008}{\emph{Fortsch. Phys.}
  {\bfseries 70} (2022) 2200008}
  [\href{https://arxiv.org/abs/2112.10783}{{\ttfamily 2112.10783}}].

\bibitem{Rudelius:2021azq}
T.~Rudelius, \emph{{Asymptotic observables and the swampland}},
  \href{https://doi.org/10.1103/PhysRevD.104.126023}{\emph{Phys. Rev. D}
  {\bfseries 104} (2021) 126023}
  [\href{https://arxiv.org/abs/2106.09026}{{\ttfamily 2106.09026}}].

\bibitem{Rudelius:2022gbz}
T.~Rudelius, \emph{{Asymptotic scalar field cosmology in string theory}},
  \href{https://doi.org/10.1007/JHEP10(2022)018}{\emph{JHEP} {\bfseries 10}
  (2022) 018} [\href{https://arxiv.org/abs/2208.08989}{{\ttfamily
  2208.08989}}].

\bibitem{Andriot:2022xjh}
D.~Andriot and L.~Horer, \emph{{(Quasi-) de Sitter solutions across dimensions
  and the TCC bound}},
  \href{https://doi.org/10.1007/JHEP01(2023)020}{\emph{JHEP} {\bfseries 01}
  (2023) 020} [\href{https://arxiv.org/abs/2208.14462}{{\ttfamily
  2208.14462}}].

\bibitem{Shiu:2023nph}
G.~Shiu, F.~Tonioni and H.~V. Tran, \emph{{Accelerating universe at the end of
  time}}, \href{https://doi.org/10.1103/PhysRevD.108.063527}{\emph{Phys. Rev.
  D} {\bfseries 108} (2023) 063527}
  [\href{https://arxiv.org/abs/2303.03418}{{\ttfamily 2303.03418}}].

\bibitem{Shiu:2023fhb}
G.~Shiu, F.~Tonioni and H.~V. Tran, \emph{{Late-time attractors and cosmic
  acceleration}},
  \href{https://doi.org/10.1103/PhysRevD.108.063528}{\emph{Phys. Rev. D}
  {\bfseries 108} (2023) 063528}
  [\href{https://arxiv.org/abs/2306.07327}{{\ttfamily 2306.07327}}].

\bibitem{Hebecker:2023qke}
A.~Hebecker, S.~Schreyer and G.~Venken, \emph{{No asymptotic acceleration
  without higher-dimensional de Sitter vacua}},
  \href{https://doi.org/10.1007/JHEP11(2023)173}{\emph{JHEP} {\bfseries 11}
  (2023) 173} [\href{https://arxiv.org/abs/2306.17213}{{\ttfamily
  2306.17213}}].

\bibitem{Freigang:2023ogu}
J.~Freigang, D.~Lust, G.-E. Nian and M.~Scalisi, \emph{{Cosmic acceleration and
  turns in the Swampland}},
  \href{https://doi.org/10.1088/1475-7516/2023/11/080}{\emph{JCAP} {\bfseries
  11} (2023) 080} [\href{https://arxiv.org/abs/2306.17217}{{\ttfamily
  2306.17217}}].

\bibitem{Andriot:2023wvg}
D.~Andriot, D.~Tsimpis and T.~Wrase, \emph{{Accelerated expansion of an open
  universe and string theory realizations}},
  \href{https://doi.org/10.1103/PhysRevD.108.123515}{\emph{Phys. Rev. D}
  {\bfseries 108} (2023) 123515}
  [\href{https://arxiv.org/abs/2309.03938}{{\ttfamily 2309.03938}}].

\bibitem{Andriot:2024jsh}
D.~Andriot, S.~Parameswaran, D.~Tsimpis, T.~Wrase and I.~Zavala,
  \emph{{Exponential Quintessence: curved, steep and stringy?}},
  \href{https://arxiv.org/abs/2405.09323}{{\ttfamily 2405.09323}}.

\bibitem{Banerjee:2018qey}
S.~Banerjee, U.~Danielsson, G.~Dibitetto, S.~Giri and M.~Schillo,
  \emph{{Emergent de Sitter Cosmology from Decaying Anti\textendash{}de Sitter
  Space}}, \href{https://doi.org/10.1103/PhysRevLett.121.261301}{\emph{Phys.
  Rev. Lett.} {\bfseries 121} (2018) 261301}
  [\href{https://arxiv.org/abs/1807.01570}{{\ttfamily 1807.01570}}].

\bibitem{Banerjee:2019fzz}
S.~Banerjee, U.~Danielsson, G.~Dibitetto, S.~Giri and M.~Schillo, \emph{{de
  Sitter Cosmology on an expanding bubble}},
  \href{https://doi.org/10.1007/JHEP10(2019)164}{\emph{JHEP} {\bfseries 10}
  (2019) 164} [\href{https://arxiv.org/abs/1907.04268}{{\ttfamily
  1907.04268}}].

\bibitem{Banerjee:2020wix}
S.~Banerjee, U.~Danielsson and S.~Giri, \emph{{Dark bubbles: decorating the
  wall}}, \href{https://doi.org/10.1007/JHEP04(2020)085}{\emph{JHEP} {\bfseries
  04} (2020) 085} [\href{https://arxiv.org/abs/2001.07433}{{\ttfamily
  2001.07433}}].

\bibitem{Banerjee:2020wov}
S.~Banerjee, U.~Danielsson and S.~Giri, \emph{{Bubble needs strings}},
  \href{https://doi.org/10.1007/JHEP03(2021)250}{\emph{JHEP} {\bfseries 21}
  (2020) 250} [\href{https://arxiv.org/abs/2009.01597}{{\ttfamily
  2009.01597}}].

\bibitem{Banerjee:2021qei}
S.~Banerjee, U.~Danielsson and S.~Giri, \emph{{Dark bubbles and black holes}},
  \href{https://doi.org/10.1007/JHEP09(2021)158}{\emph{JHEP} {\bfseries 09}
  (2021) 158} [\href{https://arxiv.org/abs/2102.02164}{{\ttfamily
  2102.02164}}].

\bibitem{Danielsson:2022fhd}
U.~Danielsson, D.~Panizo and R.~Tielemans, \emph{{Gravitational waves in dark
  bubble cosmology}},
  \href{https://doi.org/10.1103/PhysRevD.106.024002}{\emph{Phys. Rev. D}
  {\bfseries 106} (2022) 024002}
  [\href{https://arxiv.org/abs/2202.00545}{{\ttfamily 2202.00545}}].

\bibitem{Danielsson:2022lsl}
U.~Danielsson, O.~Henriksson and D.~Panizo, \emph{{Stringy realization of a
  small and positive cosmological constant in dark bubble cosmology}},
  \href{https://doi.org/10.1103/PhysRevD.107.026020}{\emph{Phys. Rev. D}
  {\bfseries 107} (2023) 026020}
  [\href{https://arxiv.org/abs/2211.10191}{{\ttfamily 2211.10191}}].

\bibitem{Danielsson:2023alz}
U.~Danielsson and D.~Panizo, \emph{{Experimental tests of dark bubble
  cosmology}}, \href{https://doi.org/10.1103/PhysRevD.109.026003}{\emph{Phys.
  Rev. D} {\bfseries 109} (2024) 026003}
  [\href{https://arxiv.org/abs/2311.14589}{{\ttfamily 2311.14589}}].

\bibitem{Basile:2023tvh}
I.~Basile, U.~Danielsson, S.~Giri and D.~Panizo, \emph{{Shedding light on dark
  bubble cosmology}},
  \href{https://doi.org/10.1007/JHEP02(2024)112}{\emph{JHEP} {\bfseries 02}
  (2024) 112} [\href{https://arxiv.org/abs/2310.15032}{{\ttfamily
  2310.15032}}].

\bibitem{Banerjee:2023uto}
S.~Banerjee, U.~Danielsson and M.~Zemsch, \emph{{The dark bubbleography}},
  \href{https://doi.org/10.1007/JHEP02(2024)102}{\emph{JHEP} {\bfseries 02}
  (2024) 102} [\href{https://arxiv.org/abs/2311.16242}{{\ttfamily
  2311.16242}}].

\bibitem{Banerjee:2021yrb}
S.~Banerjee, U.~Danielsson and S.~Giri, \emph{{Curing with Hemlock: Escaping
  the swampland using instabilities from string theory}},
  \href{https://doi.org/10.1142/S0218271821420293}{\emph{Int. J. Mod. Phys. D}
  {\bfseries 30} (2021) 2142029}
  [\href{https://arxiv.org/abs/2103.17121}{{\ttfamily 2103.17121}}].

\bibitem{Banerjee:2022ree}
S.~Banerjee, U.~Danielsson and S.~Giri, \emph{{Features of a dark energy model
  in string theory}},
  \href{https://doi.org/10.1103/PhysRevD.108.126009}{\emph{Phys. Rev. D}
  {\bfseries 108} (2023) 126009}
  [\href{https://arxiv.org/abs/2212.14004}{{\ttfamily 2212.14004}}].

\bibitem{Berglund:2022qsb}
P.~Berglund, T.~H\"ubsch and D.~Minic, \emph{{On de Sitter Spacetime and String
  Theory}}, \href{https://doi.org/10.1142/S0218271823300021}{\emph{Int. J. Mod.
  Phys. D} {\bfseries 32} (2023) 2330002}
  [\href{https://arxiv.org/abs/2212.06086}{{\ttfamily 2212.06086}}].

\bibitem{Bandos:2023yyo}
I.~Bandos, J.~J. Blanco-Pillado, K.~Sousa and M.~A. Urkiola, \emph{{Brane
  nucleation in supersymmetric models}},
  \href{https://doi.org/10.1007/JHEP10(2023)061}{\emph{JHEP} {\bfseries 10}
  (2023) 061} [\href{https://arxiv.org/abs/2306.09412}{{\ttfamily
  2306.09412}}].

\bibitem{Koga:2019yzj}
I.~Koga and Y.~Ookouchi, \emph{{Catalytic creation of baby bubble universe with
  small positive cosmological constant}},
  \href{https://doi.org/10.1007/JHEP10(2019)281}{\emph{JHEP} {\bfseries 10}
  (2019) 281} [\href{https://arxiv.org/abs/1909.03014}{{\ttfamily
  1909.03014}}].

\bibitem{Koga:2020jok}
I.~Koga and Y.~Ookouchi, \emph{{Catalytic creation of a bubble universe induced
  by quintessence in five dimensions}},
  \href{https://doi.org/10.1103/PhysRevD.104.126015}{\emph{Phys. Rev. D}
  {\bfseries 104} (2021) 126015}
  [\href{https://arxiv.org/abs/2011.07437}{{\ttfamily 2011.07437}}].

\bibitem{Koga:2022opd}
I.~Koga, N.~Oshita and K.~Ueda, \emph{{dS$_{4}$ universe emergent from
  Kerr-AdS$_{5}$ spacetime: bubble nucleation catalyzed by a black hole}},
  \href{https://doi.org/10.1007/JHEP05(2023)107}{\emph{JHEP} {\bfseries 05}
  (2023) 107} [\href{https://arxiv.org/abs/2209.05625}{{\ttfamily
  2209.05625}}].

\bibitem{Basile:2020mpt}
I.~Basile and S.~Lanza, \emph{{de Sitter in non-supersymmetric string theories:
  no-go theorems and brane-worlds}},
  \href{https://doi.org/10.1007/JHEP10(2020)108}{\emph{JHEP} {\bfseries 10}
  (2020) 108} [\href{https://arxiv.org/abs/2007.13757}{{\ttfamily
  2007.13757}}].

\bibitem{Basile:2021vxh}
I.~Basile, \emph{{Supersymmetry breaking and stability in string vacua: Brane
  dynamics, bubbles and the swampland}},
  \href{https://doi.org/10.1007/s40766-021-00024-9}{\emph{Riv. Nuovo Cim.}
  {\bfseries 44} (2021) 499}
  [\href{https://arxiv.org/abs/2107.02814}{{\ttfamily 2107.02814}}].

\bibitem{Nariai:1999}
H.~Nariai, \emph{On some static solutions of einstein's gravitational field
  equations in a spherically symmetric case},
  \href{https://doi.org/10.1023/A:1026698508110}{\emph{General Relativity and
  Gravitation} {\bfseries 31} (1999) 951}.

\bibitem{Romans:1991nq}
L.~J. Romans, \emph{{Supersymmetric, cold and lukewarm black holes in
  cosmological Einstein-Maxwell theory}},
  \href{https://doi.org/10.1016/0550-3213(92)90684-4}{\emph{Nucl. Phys. B}
  {\bfseries 383} (1992) 395}
  [\href{https://arxiv.org/abs/hep-th/9203018}{{\ttfamily hep-th/9203018}}].

\bibitem{Montero:2019ekk}
M.~Montero, T.~Van~Riet and G.~Venken, \emph{{Festina Lente: EFT Constraints
  from Charged Black Hole Evaporation in de Sitter}},
  \href{https://doi.org/10.1007/JHEP01(2020)039}{\emph{JHEP} {\bfseries 01}
  (2020) 039} [\href{https://arxiv.org/abs/1910.01648}{{\ttfamily
  1910.01648}}].

\bibitem{Hawking:1995ap}
S.~W. Hawking and S.~F. Ross, \emph{{Duality between electric and magnetic
  black holes}}, \href{https://doi.org/10.1103/PhysRevD.52.5865}{\emph{Phys.
  Rev. D} {\bfseries 52} (1995) 5865}
  [\href{https://arxiv.org/abs/hep-th/9504019}{{\ttfamily hep-th/9504019}}].

\bibitem{Bousso:1996au}
R.~Bousso and S.~W. Hawking, \emph{{Pair creation of black holes during
  inflation}}, \href{https://doi.org/10.1103/PhysRevD.54.6312}{\emph{Phys. Rev.
  D} {\bfseries 54} (1996) 6312}
  [\href{https://arxiv.org/abs/gr-qc/9606052}{{\ttfamily gr-qc/9606052}}].

\bibitem{Cai:2004eh}
R.-G. Cai, D.-W. Pang and A.~Wang, \emph{{Born-Infeld black holes in (A)dS
  spaces}}, \href{https://doi.org/10.1103/PhysRevD.70.124034}{\emph{Phys. Rev.
  D} {\bfseries 70} (2004) 124034}
  [\href{https://arxiv.org/abs/hep-th/0410158}{{\ttfamily hep-th/0410158}}].

\bibitem{Bousso:1996pn}
R.~Bousso, \emph{{Charged Nariai black holes with a dilaton}},
  \href{https://doi.org/10.1103/PhysRevD.55.3614}{\emph{Phys. Rev. D}
  {\bfseries 55} (1997) 3614}
  [\href{https://arxiv.org/abs/gr-qc/9608053}{{\ttfamily gr-qc/9608053}}].

\bibitem{Bernamonti:2007bu}
A.~Bernamonti, M.~M. Caldarelli, D.~Klemm, R.~Olea, C.~Sieg and E.~Zorzan,
  \emph{{Black strings in AdS(5)}},
  \href{https://doi.org/10.1088/1126-6708/2008/01/061}{\emph{JHEP} {\bfseries
  01} (2008) 061} [\href{https://arxiv.org/abs/0708.2402}{{\ttfamily
  0708.2402}}].

\bibitem{Klemm:2000nj}
D.~Klemm and W.~A. Sabra, \emph{{Supersymmetry of black strings in D = 5 gauged
  supergravities}},
  \href{https://doi.org/10.1103/PhysRevD.62.024003}{\emph{Phys. Rev. D}
  {\bfseries 62} (2000) 024003}
  [\href{https://arxiv.org/abs/hep-th/0001131}{{\ttfamily hep-th/0001131}}].

\bibitem{Brihaye:2007ju}
Y.~Brihaye, T.~Delsate and E.~Radu, \emph{{On the stability of AdS black
  strings}}, \href{https://doi.org/10.1016/j.physletb.2008.03.008}{\emph{Phys.
  Lett. B} {\bfseries 662} (2008) 264}
  [\href{https://arxiv.org/abs/0710.4034}{{\ttfamily 0710.4034}}].

\bibitem{Brihaye:2009dm}
Y.~Brihaye, J.~Kunz and E.~Radu, \emph{{From black strings to black holes:
  Nuttier and squashed AdS(5) solutions}},
  \href{https://doi.org/10.1088/1126-6708/2009/08/025}{\emph{JHEP} {\bfseries
  08} (2009) 025} [\href{https://arxiv.org/abs/0904.1566}{{\ttfamily
  0904.1566}}].

\bibitem{Harmark:2007md}
T.~Harmark, V.~Niarchos and N.~A. Obers, \emph{{Instabilities of black strings
  and branes}}, \href{https://doi.org/10.1088/0264-9381/24/8/R01}{\emph{Class.
  Quant. Grav.} {\bfseries 24} (2007) R1}
  [\href{https://arxiv.org/abs/hep-th/0701022}{{\ttfamily hep-th/0701022}}].

\bibitem{Mann:2006yi}
R.~B. Mann, E.~Radu and C.~Stelea, \emph{{Black string solutions with negative
  cosmological constant}},
  \href{https://doi.org/10.1088/1126-6708/2006/09/073}{\emph{JHEP} {\bfseries
  09} (2006) 073} [\href{https://arxiv.org/abs/hep-th/0604205}{{\ttfamily
  hep-th/0604205}}].

\bibitem{Chamseddine:1999xk}
A.~H. Chamseddine and W.~A. Sabra, \emph{{Magnetic strings in five-dimensional
  gauged supergravity theories}},
  \href{https://doi.org/10.1016/S0370-2693(00)00178-7}{\emph{Phys. Lett. B}
  {\bfseries 477} (2000) 329}
  [\href{https://arxiv.org/abs/hep-th/9911195}{{\ttfamily hep-th/9911195}}].

\bibitem{Israel:1966rt}
W.~Israel, \emph{{Singular hypersurfaces and thin shells in general
  relativity}}, \href{https://doi.org/10.1007/BF02710419}{\emph{Nuovo Cim. B}
  {\bfseries 44S10} (1966) 1}.

\bibitem{Danielsson:2022odq}
U.~Danielsson, V.~Van~Hemelryck and T.~Van~Riet, \emph{{Over-extremal brane
  shells from string theory?}},
  \href{https://doi.org/10.1088/1361-6382/ac96c4}{\emph{Class. Quant. Grav.}
  {\bfseries 39} (2022) 235001}
  [\href{https://arxiv.org/abs/2206.04506}{{\ttfamily 2206.04506}}].

\bibitem{Brown:1988kg}
J.~D. Brown and C.~Teitelboim, \emph{{Neutralization of the Cosmological
  Constant by Membrane Creation}},
  \href{https://doi.org/10.1016/0550-3213(88)90559-7}{\emph{Nucl. Phys. B}
  {\bfseries 297} (1988) 787}.

\bibitem{Gregory:1993vy}
R.~Gregory and R.~Laflamme, \emph{{Black strings and p-branes are unstable}},
  \href{https://doi.org/10.1103/PhysRevLett.70.2837}{\emph{Phys. Rev. Lett.}
  {\bfseries 70} (1993) 2837}
  [\href{https://arxiv.org/abs/hep-th/9301052}{{\ttfamily hep-th/9301052}}].

\bibitem{Gubser:2000ec}
S.~S. Gubser and I.~Mitra, \emph{{Instability of charged black holes in Anti-de
  Sitter space}}, {\emph{Clay Math. Proc.} {\bfseries 1} (2002) 221}
  [\href{https://arxiv.org/abs/hep-th/0009126}{{\ttfamily hep-th/0009126}}].

\bibitem{Gubser:2000mm}
S.~S. Gubser and I.~Mitra, \emph{{The Evolution of unstable black holes in
  anti-de Sitter space}},
  \href{https://doi.org/10.1088/1126-6708/2001/08/018}{\emph{JHEP} {\bfseries
  08} (2001) 018} [\href{https://arxiv.org/abs/hep-th/0011127}{{\ttfamily
  hep-th/0011127}}].

\bibitem{Friess:2005zp}
J.~J. Friess, S.~S. Gubser and I.~Mitra, \emph{{Counter-examples to the
  correlated stability conjecture}},
  \href{https://doi.org/10.1103/PhysRevD.72.104019}{\emph{Phys. Rev. D}
  {\bfseries 72} (2005) 104019}
  [\href{https://arxiv.org/abs/hep-th/0508220}{{\ttfamily hep-th/0508220}}].

\bibitem{Buchel:2010wk}
A.~Buchel and C.~Pagnutti, \emph{{Correlated stability conjecture revisited}},
  \href{https://doi.org/10.1016/j.physletb.2011.01.057}{\emph{Phys. Lett. B}
  {\bfseries 697} (2011) 168}
  [\href{https://arxiv.org/abs/1010.5748}{{\ttfamily 1010.5748}}].

\bibitem{Buchel:2011ra}
A.~Buchel and A.~Patrushev, \emph{{Can the correlated stability conjecture be
  saved?}}, \href{https://doi.org/10.1007/JHEP06(2011)090}{\emph{JHEP}
  {\bfseries 06} (2011) 090} [\href{https://arxiv.org/abs/1102.5331}{{\ttfamily
  1102.5331}}].

\bibitem{Montero:2021otb}
M.~Montero, C.~Vafa, T.~Van~Riet and G.~Venken, \emph{{The FL bound and its
  phenomenological implications}},
  \href{https://doi.org/10.1007/JHEP10(2021)009}{\emph{JHEP} {\bfseries 10}
  (2021) 009} [\href{https://arxiv.org/abs/2106.07650}{{\ttfamily
  2106.07650}}].

\end{thebibliography}\endgroup

\end{document}